\pdfoutput=1
\documentclass[twocolumn,amsmath,trackchanges]{aastex702}
\usepackage[utf8]{inputenc}
\usepackage{natbib}
\usepackage{graphicx}
\usepackage{subcaption}
\usepackage{todonotes}
\usepackage{booktabs}
\usepackage{adjustbox}

\begin{document}

\title{Constraining the Light Curves of  Highly Magnified Individual Stars at $\boldsymbol{z \approx 0.725}$}

\author[0009-0000-0378-286X]{Carter Meyerhoff}
\email{cdm4cp@umsystem.edu}
\affiliation{Department of Physics and Astronomy, University of Missouri, Columbia, MO 65211}
\affiliation{School of Earth and Space Exploration, Arizona State University, Tempe, AZ 85287-1404, USA}

\author[0000-0002-1681-0767]{Hayley Williams}
\email{hwill102@asu.edu}
\affiliation{School of Earth and Space Exploration, Arizona State University, Tempe, AZ 85287-1404, USA}

\author[0000-0002-6150-833X]{Rafael Ortiz III}
\email{rortizii@asu.edu}
\affiliation{School of Earth and Space Exploration, Arizona State University, Tempe, AZ 85287-1404, USA}

\author[orcid=0009-0007-0782-0721]{Gibson B.\ Bowling}
\email{gbbowlin@asu.edu}
\affiliation{School of Earth and Space Exploration, Arizona State University, Tempe, AZ 85287-1404, USA}
\affiliation{Department of Physics and Astronomy, University of Missouri, Columbia, MO 65211}

\author[orcid=0000-0001-8156-6281]{Rogier A. Windhorst}
\email{}
\affiliation{School of Earth and Space Exploration, Arizona State University, Tempe, AZ 85287-1404, USA}

\author[orcid=0009-0006-8766-4513]{Zach Borowiak}
\email{}
\affiliation{Department of Physics and Astronomy, University of Missouri, Columbia, MO 65211}

\author[orcid=0009-0001-7166-2453]{Hunter Sparks}
\email{}
\affiliation{Department of Physics and Astronomy, University of Missouri, Columbia, MO 65211}

\author[0000-0003-3780-6801]{Charles L. Steinhardt}
\email{}
\affiliation{Department of Physics and Astronomy, University of Missouri, Columbia, MO 65211}

\begin{abstract}

Extreme magnification by lensing galaxy clusters allows the detection of high-luminosity stars during cluster caustic transient events at cosmological distances. The Dragon Arc at $z=0.725$ is a frontier target for identifying transient events due to its high frequency of microlensing events, and so far $\gtrsim100$ of these events have been detected. The high frequency of events and high cadence of observations allow for the measurement of minimal flux variation between epochs, and therefore the magnification evolution as a source approaches or recedes from a microcaustic can be measured from light curves. In this work, we model the light curves of 10 transient events in the Dragon Arc across eight epochs from two JWST/NIRCam medium-band filters taken over four days. Injected synthetic observations demonstrate that with additional observations, individual transient light curves allow the of apparent degeneracies between stellar characteristics and properties of motion. 

\end{abstract}

\section{Introduction}
Microlensing transient events occur when point-like background sources, e.g., individual stars, move near the critical curves of galaxy cluster-scale gravitational lenses and experience brief periods of extreme magnification due to microlensing from an intracluster star \citep{Kaurov2019,Yan2023,Williams2026}. Microlensing contributes magnification factors on the scale of several hundreds to thousands \citep{Venumadhav2017,Diego2018}, and can significantly contribute to the overall frequency of transient events within a lensed galaxy. The galaxy with the highest known frequency of microlensing transient events is the ``Dragon'' Arc \citep[$z=0.725$;][]{Soucail1988}. The Dragon arc's low redshift and high surface brightness, along with the multiple critical curve crossings across the arc make it an ideal site for observing lensed stars and other point sources, allowing for the identification of $\gtrsim100$ transient events over multiple epochs \citep{Fudamoto2025,palencia2026}.  

Previous analyses of microlensing transient events have typically been limited to information that can be extracted from individual epoch identification, as higher cadence observations are often stacked to increase SNR and improve detectability. While the stacking process makes transients easier to detect, it makes it difficult to evaluate motion and stellar parameters of transient events, and restricts further study to event tracking and population analysis. On shorter time scales, multi-epoch observations enable the observation of transient flux variation and modeling of stellar kinematics \citep{Williams2026}. Flux evolution should follow a predictable light curve as a source approaches or recedes from the caustic, as determined from \citet{MiraldaEscude1991}. These light curves give valuable insight into microlensing events within the Dragon arc, and the determination of transient properties could constrain micro and macrolens interactions within strongly lensed galaxies, and allow the observation of compact dark matter (DM) interactions or density fluctuations associated with Fuzzy Dark Matter (FDM)\citep{Oguri2018,Diego2018,Dai2018,Diego2024,Palencia2024,Broadhurst2025,Ji2025,Muller2025,Croon2026}.


In this paper, we model the light curves of 10 transient events in the Dragon arc using 8 medium-band JWST/NIRCam observations spanning 4 days. In addition, we inject additional simulated observations to test the necessary cadence and baseline to constrain stellar velocities and radii. Analyzing light curves before and after injected observation dates demonstrates the impact of high-cadence observation when constraining stellar properties.

This paper is organized as follows. Section \ref{sec:data} describes the publicly available JWST/NIRCam data used. Section \ref{sec:detections} describes the methodology utilized for PSF reconstruction, residual analysis, and transient detection. Section \ref{sec:fluxvar} describes the observation of flux variation between epochs for each detected transient event. Section \ref{sec:lightcurves} describes results of light curve fitting and the impact of additional observation dates on constrained parameters. Section \ref{sec:results} discusses the main results of this paper and potential improvements for this analysis, along with future work.

\section{Data}
\label{sec:data}

The data was retrieved from publicly available JWST/NIRCam imaging taken by the programs listed in Table~\ref{tab:table1}, GO-3538 \citep[PI Iani;][]{Iani2023} and GO-2883 \citep[PI Sun;][]{sun2023}. The raw exposures were downloaded and reduced using version 1.20.2 of the JWST Science Calibration Pipeline (\texttt{CALWEBB}; STScI) \citep{Bushouse2017}, with calibration reference files selected via the Calibration Reference Data System context \texttt{jwst\_1464.pmap}.
To mitigate \texttt{JUMP} artifacts
during stage 1, we customized the standard pipeline and its three stages following the PEARLS
\citep{Windhorst_2023} prescriptions. We removed both $1/f$ noise and ``wisps'' (i.e., straylight artifacts on the short wavelength-channel detectors) by applying the \texttt{JumProPe} \citep{Robotham2023, DSilva2025} suite on the calibrated exposure files from stage 2. Astrometry of the image mosaic was corrected to the WCS alignment of GAIA DR3 \citep{Gaia_2023}. 
Mosaics were drizzled with a pixel scale of 0\farcs{03}/pix grid across the filters, all in the north-up direction. The F182M and F210M mosaics are those used in this study. Images were aligned onto the same WCS grid with pixel-by-pixel matching from a modified stage 3 JWST pipeline.

The GO-3538 program was utilized for this analysis due to its high cadence observations spanning 4 days across 2 filters, F182M and F210M. These high cadence observations have similar PA's, and allow light curves to be measured from flux variation between epochs. The GO-2883 program contained images with higher exposure and S/N ratios for both F182M and F210M at a later epoch, and therefore provided a strong reference for residual analysis.

All of the {\it JWST} data used in this paper can be found in MAST: \dataset[10.17909/cbaj-0c82]{http://dx.doi.org/10.17909/cbaj-0c82}.

\begin{table}[b]
\centering
\caption{Summary of JWST exposure times organized by filter and epoch, for missions GO-2883 and GO-3538.}

\begin{adjustbox}{width=1.25\columnwidth, margin=-2.2cm 0cm 0cm 0cm}
\begin{tabular}{|l|c|c|c|r|}
\hline
\textbf{Program} & \textbf{Filter} & \textbf{Date} & \textbf{UTC} & \textbf{Exposure (s)} \\ \hline
GO-3538 & F182M & 2023-12-19 & 08:06:04 & 4638   \\ \hline
GO-3538 & F182M & 2023-12-19 & 16:15:41 & 5067  \\ \hline
GO-3538 & F182M & 2023-12-20 & 12:28:34 & 4638   \\ \hline
GO-3538 & F182M & 2023-12-22 & 19:27:19 & 4745  \\ \hline
GO-3538 & F210M & 2023-12-19 & 10:30:03 & 4638   \\ \hline
GO-3538 & F210M & 2023-12-19 & 12:57:47 & 5067  \\ \hline
GO-3538 & F210M & 2023-12-20 & 23:06:25 & 4638   \\ \hline
GO-3538 & F210M & 2023-12-22 & 17:21:58 & 5067  \\ \hline
GO-2883 & F182M & 2024-07-31 & 11:02:11 & 9276  \\ \hline
GO-2883 & F210M & 2024-07-24 & 23:22:28 & 9276  \\ \hline
\end{tabular}
\end{adjustbox}
\label{tab:table1}
\end{table}

\section{Transient Detection and Photometry}
\label{sec:detections}
The goal of this paper is to observe flux variation and constrain the light curves of transient events present within the Dragon Arc from multi-epoch measurements, and extract physical parameters. Since multi-epoch observation is only possible for transient events present in December 2023, transients must be identified from subtraction methods between each epoch present in December and its filters' corresponding reference epoch. Transient event detection for these epochs relies on stars whose differential magnification exceeds a threshold above the noise level and whose intrinsic brightness, when combined with the lensing magnification, makes detection possible given the observational depth of the reference July epoch. Data must be prepared properly before subtraction to reduce the number of false transient detections and improve source flux measurements.


In this section, we first describe PSF construction for residual analysis and transient identification. We then describe the methods of identifying transient sources within the Dragon Arc. We then describe a source of intriguing nature.

\subsection{PSFs and Residual Analysis}

Proper analysis of potential transient candidates requires differential analysis between epochs. While epochs from Program GO-3538 was designed to have matching PAs within their respective filters, Program GO-2883 was done with a different PA. To correct for this PSF mismatch, images were first grouped by filter, with 5 epochs per image. PSFs were then generated for each epoch using \texttt{photutils} \citep{photutils}, and each December PSF was convolved with the reference July PSF from the GO-2883 mission. This convolution was then applied to the July image for each December image, producing 4 PSF-matched July images for both F182M and F210M. 

\begin{figure*}[htbp]
    \centering
    \includegraphics[width=0.95\textwidth]{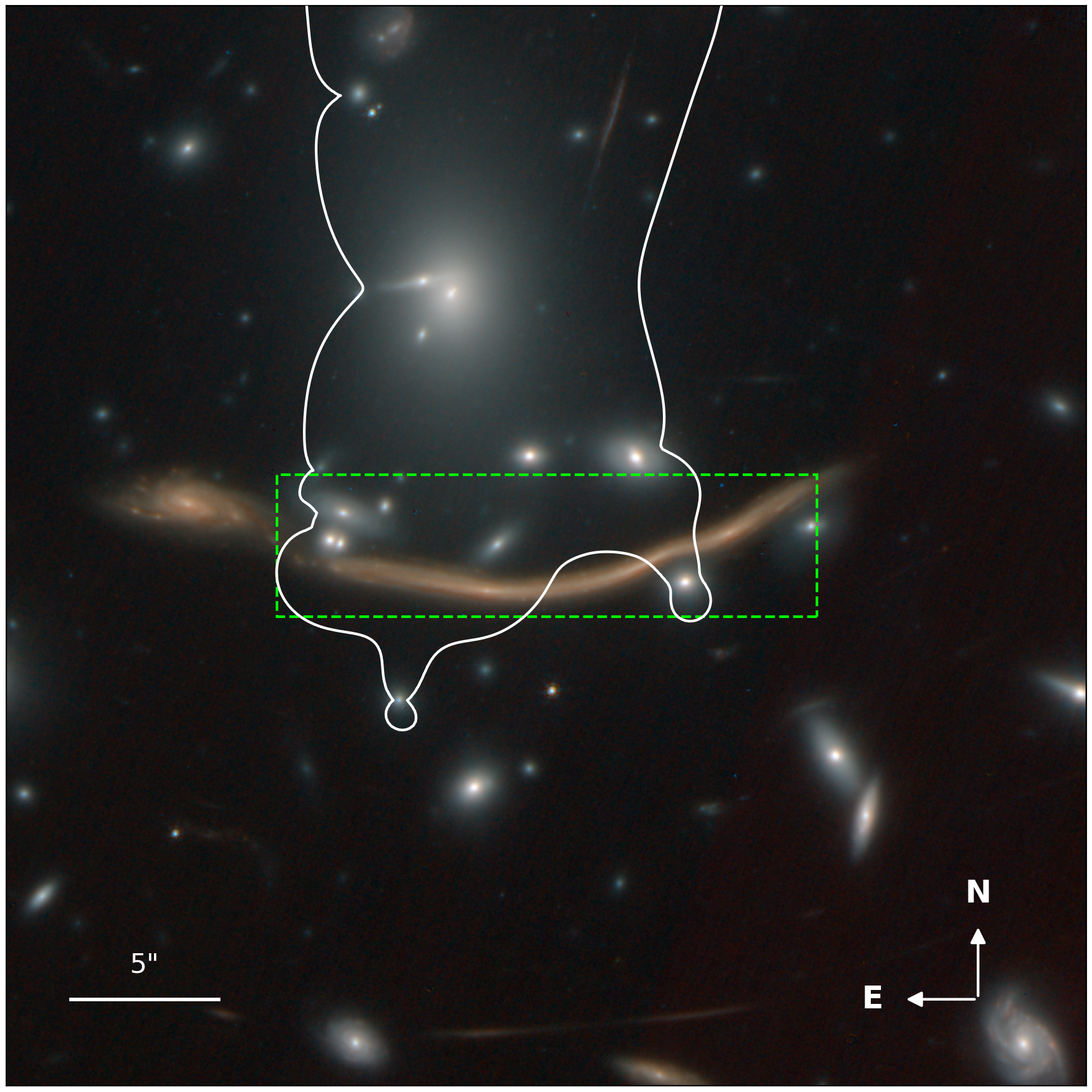}
    \caption{Cutout region of Abell 370 ($54''\times54''$) centered on the Dragon arc, with overlaid cutout region ($5''\times18''$) used for DAOStarFinder transient event location. The color rendering was made in \texttt{Trilogy}\footnote{\url{https://www.stsci.edu/~dcoe/trilogy/Intro.html}} \citep{Coe2012} using F182M, F335M, and F480M NIRCam images in blue, orange, and red channels, respectively. Overlaid are the critical curves from the lens model for Abell 370 from \citet{Niemiec2023}, the thick white line indicates regions of maximum magnification.}
    \label{fig:fulldragon}
\end{figure*}

After PSF characterization of each epoch, each July image for both F182M and F210M was aligned with their corresponding December image, and small subpixel corrections were made. These images were then subtracted, producing 8 residual images that allow for the identification of transients present from December 19 to December 23. Notably, dark spots in residual images were excluded from analysis, as they signify transients that are present in the July 2024 reference dates, and it is impossible to view variation with only one July epoch for each filter. 

We detected point sources in subtraction images using \texttt{DAOStarFinder}, a \texttt{photutils} package that parses astronomical images for local density peaks that exceed a defined background level \citep{Stetson1987}. While bright spots are easy to visually observe using difference images, \texttt{DAOStarFinder} was utilized to confirm potential transient sources in combination along with visual confirmation to ensure robustness of results. 

\begin{figure*}[h]
    \centering
    \includegraphics[width=1\textwidth]{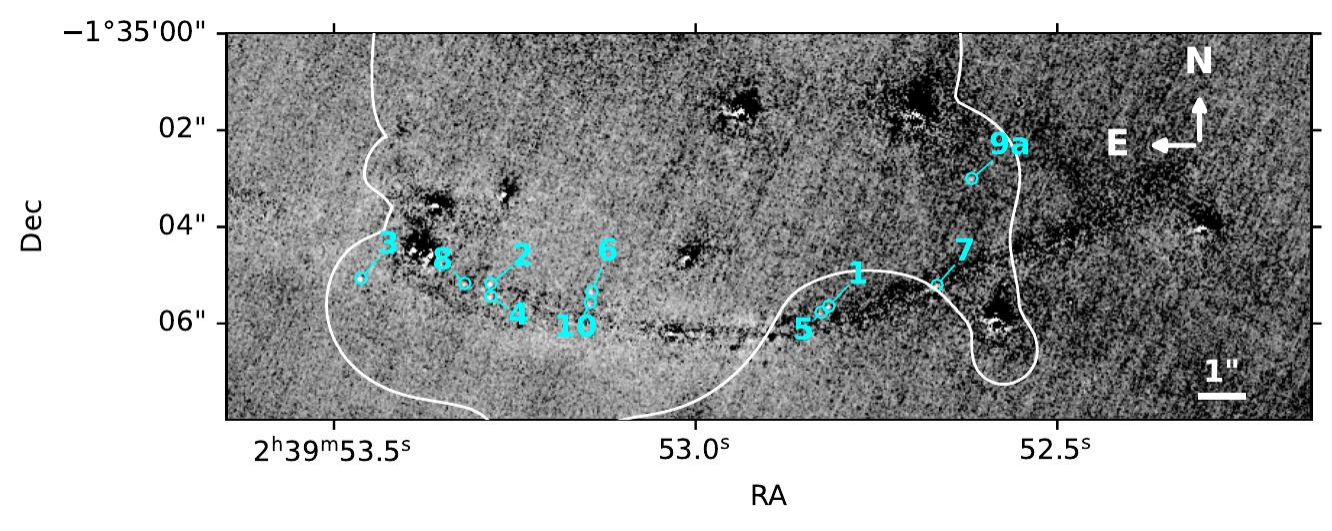}
    \caption{Visually confirmed sources identified by \texttt{DAOStarFinder}. Detections within 0.1 pixel of eachother were classified as one unique source. Source 9 is denoted as ``9a'' to indicate unique nature and absence from the \citet{palencia2026} catalog. Overlaid are the critical curves from the lens model for Abell 370 from \citet{Niemiec2023}, the thick white line indicates points of max magnification.}
    \label{fig:sources}
\end{figure*}

To limit false detections, regions of visible bright compact sources were excluded. The median background level was estimated using a sigma-clip level of 3$\sigma$, which removed background sources within 3$\sigma$ of the detection region, and sources were determined using a FWHM of 3 pixels. Detections were contained within a 5" by 18" region around the Dragon arc, centered at RA:$2^{\text{h}} 39^{\text{m}} 52.9^{\text{s}}$, and Dec:$-1^\circ 35' 4.7''$, shown in Figure~\ref{fig:fulldragon}. Detections below the 5$\sigma$ detection threshold were excluded from analysis. The minimum source separation was set to 0\farcs1. Transient detections were then visually confirmed, and false detections were excluded based on lack of visible pixel variation at the site of a detection. 10 confirmed sources are shown in Figure~\ref{fig:sources}, and a list of detections with filters and corresponding $\sigma$ values is shown in Table \ref{tab:sigmadetections}.

\begin{figure*}[htbp]
    \centering
    \includegraphics[width=1.0\textwidth]{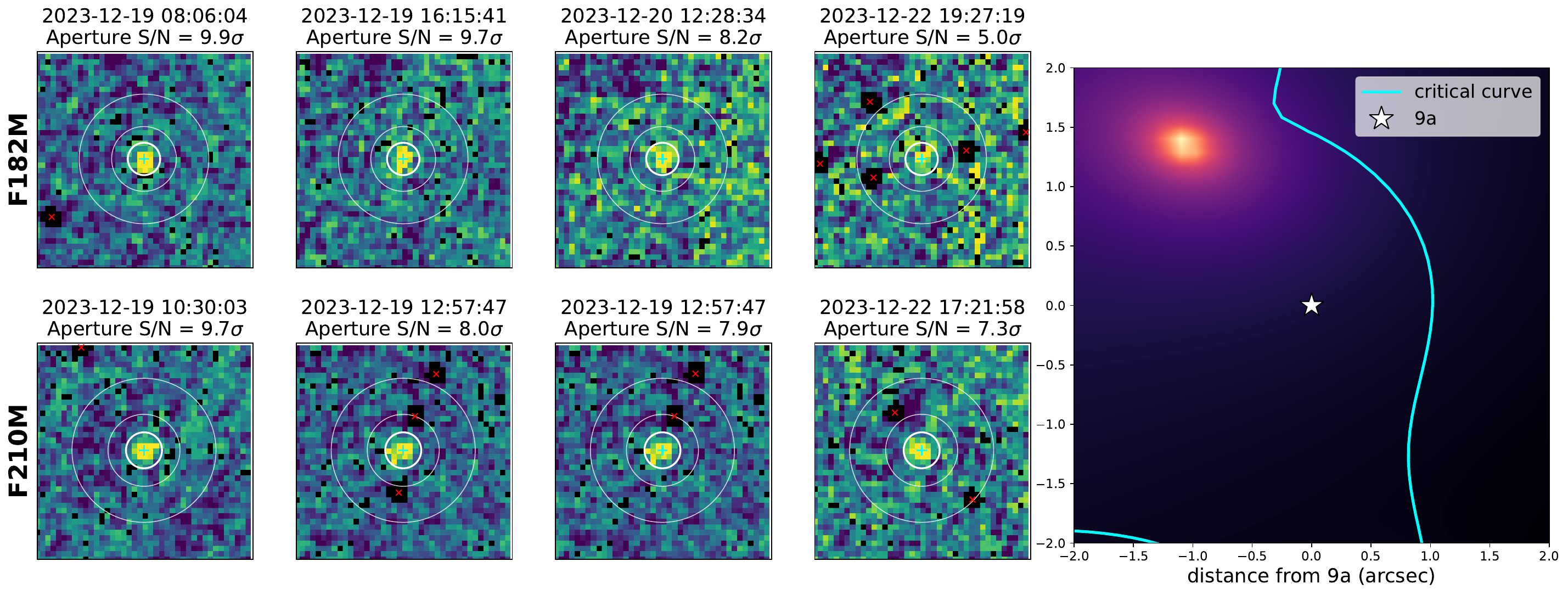}
    \caption{Individual detections of source 9a per epoch with respective S/N values. Visualization of source 9a's distance from the critical curves from \citet{Niemiec2023} is shown in the image on the right.}
    \label{fig:9aDetections}
\end{figure*}

\begin{table}[h!]
\centering
\caption{The highest $\sigma$ detection from \texttt{DAOStarFinder} per transient across all GO-2883 images subtracted from GO-3538 images. The  brightest date detection is the date with the highest recorded $\sigma$ value. Detections were visually confirmed before further analysis. Transients found with \texttt{DAOStarFinder} no detections above a $5\sigma$ threshold were excluded from analysis.}
\begin{tabular}{|l|c|c|r|}
\hline
\textbf{Source} & \textbf{F182M $\sigma$} & \textbf{F210M $\sigma$} & \textbf{Brightest Date}\\ \hline
1 & 7.99 & 11.17 & 2023-12-22 \\ \hline
2 & 10.44 & 12.30 & 2023-12-22 \\ \hline
3 & 13.94 & 12.30 & 2023-12-19 \\ \hline
4 & 5.54 & 6.45 & 2023-12-20 \\ \hline
5 & 4.60 & 6.41 & 2023-12-22 \\ \hline
6 & 5.27 & 6.30 & 2023-12-22 \\ \hline
7 & 6.23 & 5.99 & 2023-12-22 \\ \hline
8 & 6.07 & 6.79 & 2023-12-22 \\ \hline
9a & 7.32 & 4.86 & 2023-12-19 \\ \hline
10 & 6.62 & 5.74 & 2023-12-22 \\ \hline
\end{tabular}
\label{tab:sigmadetections}
\end{table}

\subsection{Aperture Photometry Measurements}

Point source photometry measurements were taken from aperture size in the F182M filter as 0\farcs090 and aperture size in the F210M filter as 0\farcs100, each containing 68.6\% of the enclosed energy \citep{Perrin2014, Rigby2023}.  To identify and mask nearby contaminating sources prior to local background estimation before detection, we ran \texttt{DAOStarFinder} on each cutout with a Gaussian kernel FWHM of 1.5 pixels and a detection threshold of 3$\sigma$ above the locally estimated (sigma-clipped) background. Furthermore, a separate upper sigma clipping pass masks pixels 2$\sigma$ above the local clipped median. Local background was then estimated from flux measurements of regions between an inner radii of 0\farcs180 and outer radii 0\farcs360 for F182M, and inner radii of 0\farcs200 and outer radii 0\farcs400" for F210M. Background flux was then subtracted from net flux measurements within the apertures to obtain final flux measurements for each source. Net flux uncertainty is added in quadrature from individual uncertainty of aperture flux and local annuli. This process was done independently for each of the 10 transient events.

\subsection{Source 9a}

We compared our detections with those found in a transient search by \citet{palencia2026}, and 9 of the 10 sources found in this paper were also present in the catalog. The source not present in the catalog is marked with "a" to denote its absence. Based on the results of the \citet{palencia2026} paper, this source is likely excluded due to distance from the Dragon arc itself, which is $\approx1.65$ arcseconds. This intriguing source is present in all F182M and F210M in December (with visual confirmation), with $\geq5\sigma$ observations in the F182M filter. This source was also visible in the F300M observation of the Dragon arc on Dec 19th, 2023, at 08:36:40. 

This source acts differently than typical transient sources in the Dragon arc due to its distance from the arc while also being near the critical curve for z = 0.725, and lacking a visible foreground galaxy. These factors could imply that the Dragon arc has a low-luminosity satellite. Another possible explanation could be that there exists a faint galaxy at the site of 9a, and a microlensing event within has magnified the source's luminosity to a visible point. Due to ambiguity, the source was kept for current analysis and treated as a transient source for the purposes of this paper. A full analysis of the nature of this source is left to future work.


\section{Flux variation}
\label{sec:fluxvar}

Following photometric measurements, peak-to-peak variability across epochs was determined for each filter, as shown in Table~\ref{tab:fluxsig}. Any observed flux variation between epochs is predicted to follow the evolution of magnification in light curves, drawn from Equation~\ref{eq:A} by \citet{MiraldaEscude1991}, where A is current magnification compared to an initial magnification $A_0$, $l_0$ is initial distance from the caustic, $R$ is stellar radius.

\begin{equation}
    A = \frac{A_0}{R^{1/2}}F \left( \frac{l_0}{R}-1 \right),
\label{eq:A}
\end{equation}
where $F(y_0)$ is given by
\begin{equation}
F(y_0) = \begin{cases} \displaystyle 
\frac{2}{\pi}\int_{-y_0}^{2}\mathrm{d}y\, \left[\frac{y(2-y)}{y+y_0}\right]^{1/2}&  (y_0 < 0), \\ 
\displaystyle \frac{2}{\pi}\int_{0}^{2}\mathrm{d}y \, \left[ \frac{y(2-y)}{y+y_0} \right]^{1/2} &  (y_0 > 0 ).
\end{cases}
\label{eq:y0}
\end{equation}  

\begin{figure}[htbp]
    \centering
    \includegraphics[width=0.45\textwidth]{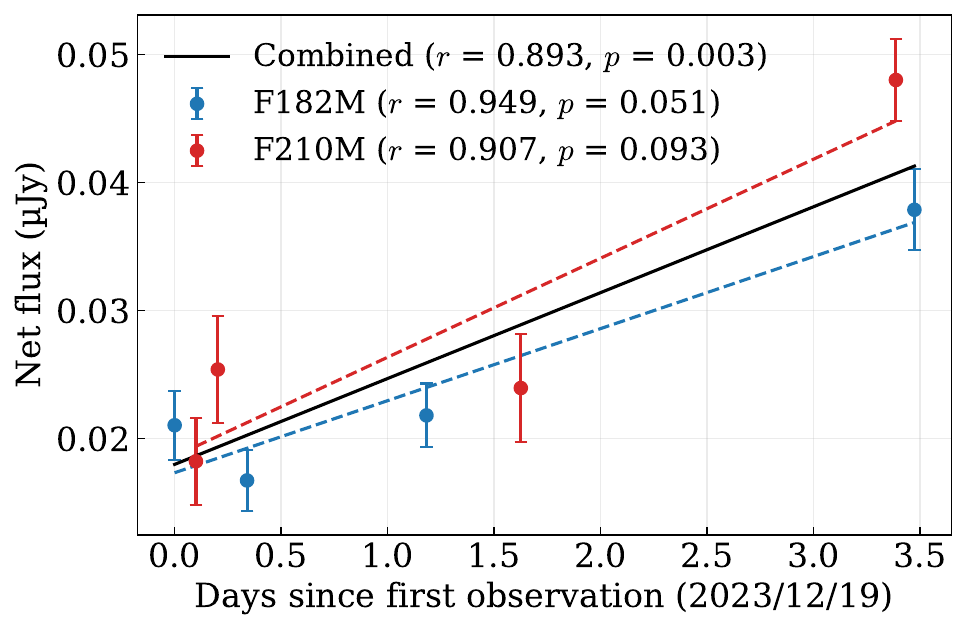}
    \caption{Net flux values were determined from raw flux values after background subtraction for Source 1 extracted from residual images. Pearson-$r$ values demonstrate evidence of variation between epochs, which is strengthened when filters are combined.}
    \label{fig:fluxvar}
\end{figure}

To determine whether enough variation is present to reasonably extract light curves from limited data, flux variation between epochs must be sufficiently significant, as sources with lower variability are much more difficult to properly constrain. The flux variation for Source 1 in F182M, F210M is shown in Figure~\ref{fig:fluxvar}.


\begin{table}[h!]
\centering
\caption{Significance values of the peak-to-peak variability for each source per filter.}
\label{tab:fluxsig}
\makebox[\columnwidth][c]{%
\begin{tabular}{|c|c|c|}
\hline
\textbf{Source} & \textbf{F182M $\sigma$} & \textbf{F210M $\sigma$} \\ \hline
1 & 5.32$\sigma$ & 6.57$\sigma$ \\ \hline
2 & 1.29$\sigma$ & 1.04$\sigma$\\ \hline
3 & 1.22$\sigma$ & 1.39$\sigma$\\ \hline
4 & 2.78$\sigma$ & 1.72$\sigma$\\ \hline
5 & 1.79$\sigma$ & 3.12$\sigma$\\ \hline
6 & 2.37$\sigma$ & 1.11$\sigma$\\ \hline
7 & 4.44$\sigma$ & 3.06$\sigma$\\ \hline
8 & 2.38$\sigma$ & 0.90$\sigma$\\ \hline
9a & 2.70$\sigma$ & 2.17$\sigma$\\ \hline
10 & 3.54$\sigma$ & 2.15$\sigma$\\ \hline
\end{tabular}%
}
\end{table}

From current observations, Source 1 is the only source with observations above $5\sigma$, or observations near the peak of the light curve. From this study, approximately 10\% of detectable transients will be close enough to the microlensing magnification peak to detect $>5\sigma$ flux variation over the course of a few days. In all, 4 out of 10 transient events have observable flux variation above $3\sigma$. This high percentage of variable transient events within a brief observation period implies that additional observations  would significantly increase the variability of these events, creating a larger population of high variability transient events. 

\section{Light Curve Modelling}
\label{sec:lightcurves}
\subsection{MCMC Fitting}

To constrain the light curves of the detected transients, modified equations from \citet{MiraldaEscude1991} were used to solve for a ratio between two magnifications ${\mu(t)}/{\mu(t_0)}$, where $t_0$ is the initial observation time. Equation~\eqref{eq:magnratio} is the modified equation to solve for magnification ratio, with $F(y_0)$ defined from Equation~\eqref{eq:y0}. We parameterize the light curves in terms of {\it relative} magnification as a function of time, rather than the absolute magnification, since we do not know the intrinsic luminosity of each source. Since relative magnification is used instead of exact magnification, for each source, the earliest positive magnification value was used as a reference epoch and set to 1.0. 

\begin{equation}
\frac{\mu(t)}{\mu(t_0)}=\frac{F\left(\frac{l_0+v_{\perp}(t-t_0)}{R}-1\right)}{F\left(\frac{l_0}{R}-1\right)}
\label{eq:magnratio}
\end{equation}

The value for $t-t_0$ used the first date with positive flux measurement in each set of measurements as $t_0$, and therefore the magnification ratio for the first data points for each filter was set to $1.0$. To reliably constrain parameters with assumedly uniform distributions, we used Markov Chain Monte Carlo (MCMC) simulations, as MCMC without priors provides reliable distributions of samples, and robustly determines convergence of parameters. The \texttt{DEMetropolisZ} sampler from \texttt{pymc} \citep{pymc2023} was utilized for MCMC simulations, and combined F182M and F210M fit were ran with three free parameters: $l_0$ in units of $\rm R_{\odot}$, $v_{\perp}$ in units of km/s, and $R$ in units of $\rm R_{\odot}$. Ten preset guesses within each parameters bounds were applied ot every source. From these guesses, the source with the lowest reduced $\chi^2$ value was used as the starting point for MCMC simulations. For each free parameter, we applied uniform priors, since no parameter bias can be assumed, with the following bounds: $-3000 ~\mathrm{R_\odot} < l_0 < 3000~\mathrm{R_\odot}$, $1.0~\mathrm{R_\odot} < R < 2000~\mathrm{R_\odot}$, and $\rm -3000~km~s^{-1} < v_\perp < 3000 km~s^{-1}$. The upper bound for the stellar radius was chosen based on the theoretical upper limit for the radius of a red supergiant \citep{Levesque2005}. The bounds for velocity were determined based on works by \citet{Angus2007, Milos2007, Sandor2013, Wilde2026}. The initial distance from the caustic was left relatively unbounded due to lack of relevant physical limits. Each simulation had 4 chains with 4000 draws each, resulting in 16000 posterior samples per fit. Pure best fits were drawn from the posterior sample with the lowest $\chi^2$ value. Errors in the best fit were drawn from the 68\% confidence interval of MCMC chain results.

Resulting best-fit light curves were able to predict how magnification changes over time, but individual parameters have larger error budgets. Apparent degeneracies between parameters of higher variability sources is easier to visualize in source corner plots than lower variability sources, but MCMC simulations struggle to determine direction of transient motion, which can lead to bimodal solutions with similar peak values, suggesting a common magnitude but different direction relative to the caustic. The corner plot for the highest variability source 1 is shown in Figure~\ref{fig:corners}.

\subsection{Injected Observations}

Limited number of observations such as the epochs used in this analysis often cause large uncertainties. As such, individual parameter uncertainties are large. In order to determine if measurements could be improved with additional observing dates, we tested the effects of additional epochs before and after current observations on our three free parameters. Synthetic observations at dates spaced outside the real observations were drawn randomly from the 68\% confidence interval and injected into the set of light curve-constraining fluxes with added noise, where added noise is the median uncertainty from each measurement within the same filter. Errors were modeled to reflect similar exposure times as real observations. The shape of the curve was found to be best determined when there were observations present near the peak of the curve, and when there were observations long before and long after the peak. Therefore, sources were injected at intervals reflecting a geometric time series, to attempt to most strongly constrain the curves. Injections were spaced geometrically using real observation dates up to 32 days on each side, to provide constraints outside of the light curves peaks. While this analysis is possible with measurements from only one filter, and there would be no significant shift in results, injected dates were alternated between F182M and F210M filters to maintain consistency with real observation dates.

After fitting, the resulting best fit parameters based on reduced $\chi^2$ value was recorded, with uncertainties based on new 16th and 84th percentile values. Resulting parameter values and errors, along with reduced $\chi^2$ values for fits before and after injections are shown in the Appendix.

\section{Discussion}

\label{sec:results}

When sources are far from crossing the microcaustic, we expect slower magnification variation with time. Sources near the microcaustic experience much higher rates of variation, therefore we only expect to see significant variation in sources which happen to be near their respective microcaustic during the time of observations. Its important to note that caustic models are smoothed idealizations of what caustics actually look like, making it difficult to distinguish variability fuzziness from real light curve variation. Therefore, results below $3\sigma$ are difficult to distinguish as real variation. Variation above $3\sigma$ is only present in 4 of the 10 present transient sources due to the short time frame of observation.  Other sources have minimal flux variation, indicating that the source was relatively far away from the microcaustic during observations. Due to the rapid evolving nature of transients, these transients are predicted to experience higher variability either earlier or later than the available observation period. 

\begin{figure*}[htbp]
    \centering
    \includegraphics[width=1.0\textwidth]{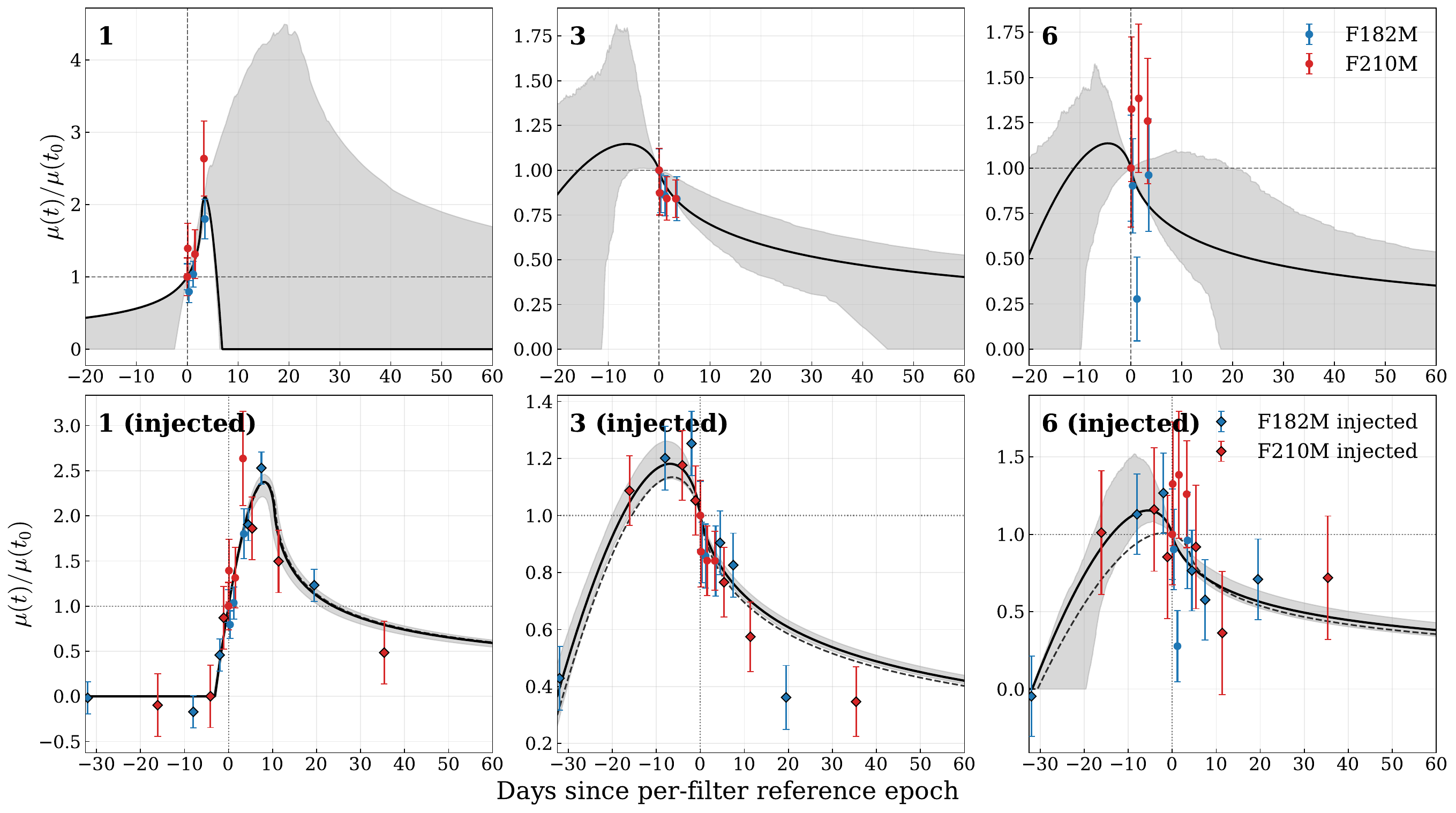}
    \caption{Source 1 (left), 3 (middle), and 6 (right) before (top) and after (bottom) injected observation dates. The solid black line is the best-fit model for the data, and the dashed line present in the bottom row is the previous best-fit model where injected observation dates were drawn from. The horizontal and vertical lines are centered on the reference epoch used for each fit. The shaded region contains the fits between the 16th and 84th percentiles, demonstrating the range of possible fits given the limited data currently available. The visualization of the light curves contains an error bar for the first source in each filter, which was retained for viewing purposes only, and does not affect the fit. Injected models mock observed data, and it can be observed that injected dates constrain the shape of the light curve much more robustly.  For Source 1, the best-fit shape inverts transient direction once more dates are observed, reinforcing the uncertainty provided with minimal epochs. }
    \label{fig:timelines}
\end{figure*}

\begin{figure*}[htbp]
    \centering
    \begin{subfigure}[b]{0.6\textwidth}
        \centering
        \includegraphics[width=\linewidth]{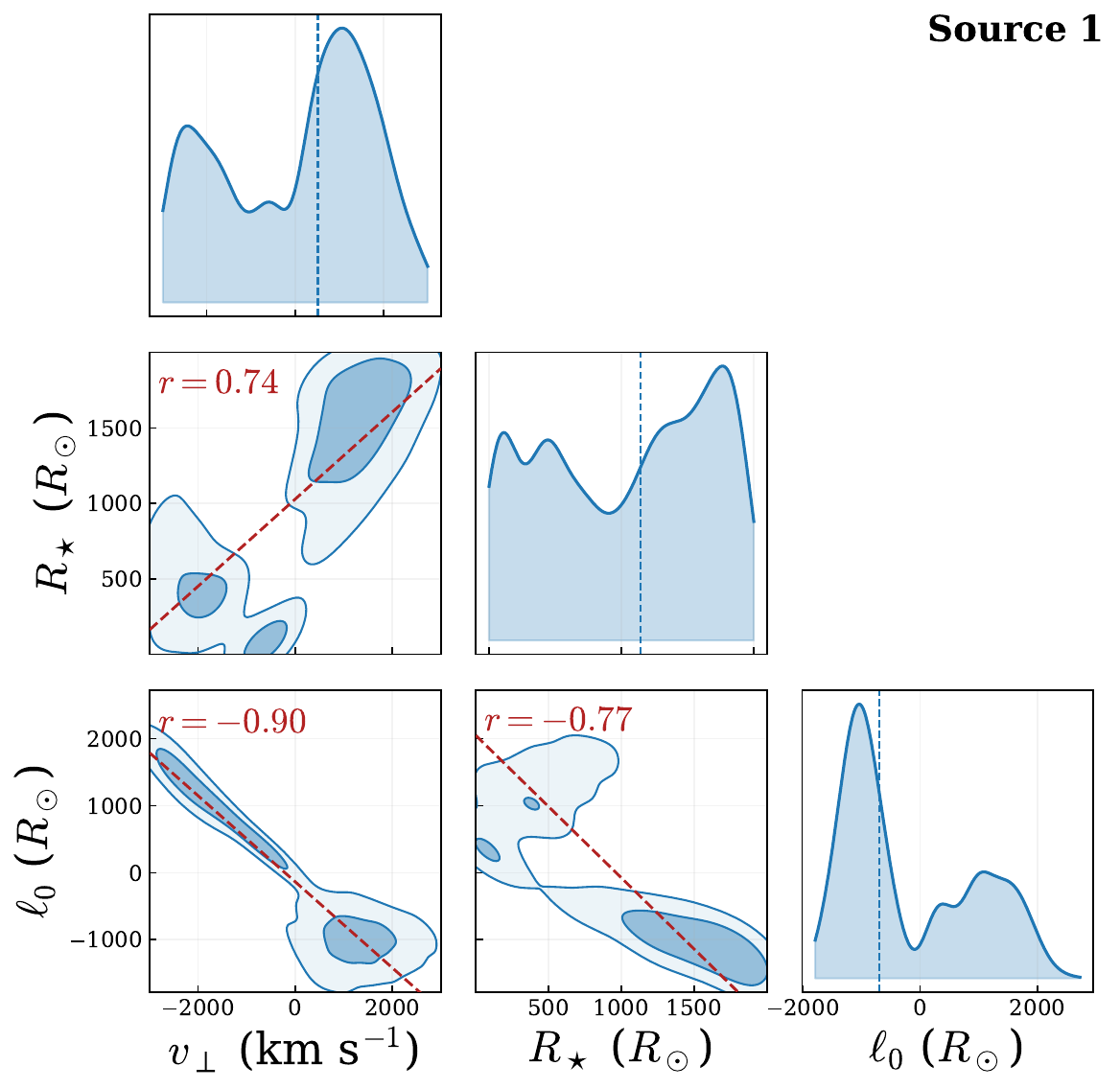}
    \end{subfigure}
    \par\bigskip
    \begin{subfigure}[b]{0.6\textwidth}
        \centering
        \includegraphics[width=\linewidth]{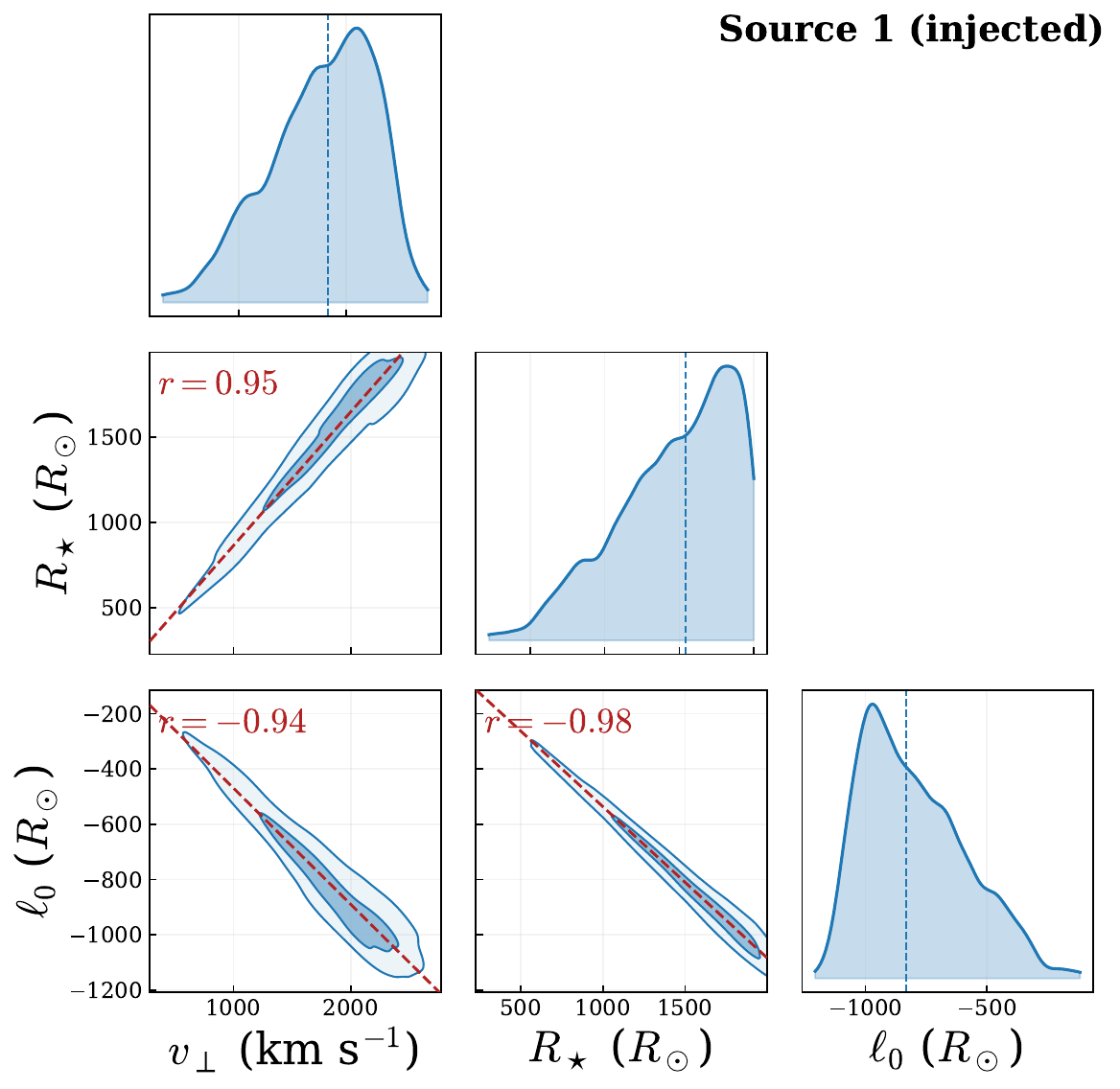}
    \end{subfigure}
    \caption{Corner plots for Source 1 referenced in Figure~\ref{fig:timelines}. Correlation between constrained parameters is difficult to determine from current observations (top). When additional epochs injected (bottom), parameters are constrained to much smaller ranges, and direct correlation becomes robust.}
    \label{fig:corners}
\end{figure*}

MCMC sampling was utilized to model said light curves of each transient as a function of stellar radius, distance from the microlensing caustic at time 0, and projected velocity with respect to the microcaustic, as shown in Equation \ref{eq:magnratio}. Several of these fits are shown in Figure \ref{fig:timelines}, with predicted best-fit evolution based on reduced chi-squared value, along with the 68\% confidence interval, drawn from fits between the 16th and 84th percentile. To better visualize degeneracies between parameters, best-fit Pearson-$r$ slopes are shown for each set of parameters in Figure~\ref{fig:corners}. 

For the 6 sources with minimal flux variation between epochs ($<3\sigma$ peak-to-peak variability), the uncertainties on the parameters from the MCMC fitting are large, as expected for measurements far from the peak of the light curve. Correlation between parameters can be robustly observed for sources with higher peak-to-peak variability, denoting an apparent degeneracy between parameters. The highest variability source shown in Figure~\ref{fig:corners} demonstrates relationships between parameters before injections, with pearson-$r$ values approaching 1. These apparent degeneracies are strengthened with injected observations, as all 3 parameters essentially possess a one-to-one correlation. Other higher-variability sources match this behavior, demonstrating the importance of observing transient events near the peak of their respective light curves. This behavior can also be observed in lower peak-to-peak variability sources with additional observations, as shown by the significant decrease in reduced $\chi^2$ value after injections. Thus, additional observations would not just increase the likelihood of observing a transient near its peak, but also improve light curve predictions for lower variability sources.

While difficult to constrain individual parameters, it is possible to identify degeneracies between parameters of the light curves of transients with increased observation time and brief high cadence observations. With enough observations, these degeneracies become almost covariant, providing insight on how each parameter interacts with each other. If able to determine one parameter robustly, the other parameters could be calculated. For example, stellar radii could be constrained from photometric measurements that constrain the surface gravity of a stellar transient \citep{Gray2019}. Current observations of the Dragon Arc lack the information to properly determine a constraint independently, additional observations across ~6 filters would constrain temperature, and therefore could be used to put an upper bound on radius based on stellar isochrones. Using this bound, the stellar radius could then be used to extract the transverse velocity and initial caustic distance from degenerate relationships shown in this paper.

Over the brief observation period in December 2023, it was possible to identify 4 sources with $>3\sigma$ peak-to-peak variability, and 1 source with $>5\sigma$ peak-to-peak variability in at least one filter. Based on the known high frequency of events within the Dragon Arc \citep{Fudamoto2025,palencia2026}, expanding the time frame of observation alone would not only allow the observation of variability in transient sources, but would also expand extensively upon current catalogs. While current limited observations allow for the viewing of flux variation between epochs, higher cadence observations would increase robustness of fitted light curves, which in turn would provide information about gravitational interactions of galaxy clusters with other massive objects. Understanding their motion and properties would also provide insight on the DM and FDM interactions \citep{Oguri2018,Diego2018,Dai2018,Diego2024,Palencia2024,Broadhurst2025,Ji2025,Muller2025,Croon2026}, help contribute to solving the IMF problem \citep{gabrielli2024}, and provide insight on stellar sources at cosmological distances. Further observations of the Dragon Arc would not only be beneficial to current studies, but would provide an opening to new transient studies that were previously impossible.

\section*{Acknowledgments}
This work is based on observations made with the NASA/ESA/CSA James Webb Space Telescope. The data was obtained from the Mikulski Archive for Space Telescopes at the Space Telescope Science Institute, which is operated by the Association of Universities for Research in Astronomy, Inc., under NASA contract NAS 5-03127 for JWST. These observations are associated with programs GO-2883 \citep[PI Sun;][]{sun2023} and GO-3538 \citep[PI Iani; ][]{Iani2023}.

This research made use of \texttt{photutils}, an Astropy package for detection and photometry of astronomical sources \citep{photutils}, and pymc \citep{pymc2023} for MCMC sampling.

Special thanks to Charles Steinhardt for funding Carter Meyerhoff.

\clearpage
\bibliographystyle{aasjournal}
\bibliography{references}

\begin{thebibliography}{}
\expandafter\ifx\csname natexlab\endcsname\relax\def\natexlab#1{#1}\fi
\providecommand{\url}[1]{\href{#1}{#1}}
\providecommand{\dodoi}[1]{doi:~\href{http://doi.org/#1}{\nolinkurl{#1}}}
\providecommand{\doeprint}[1]{\href{http://ascl.net/#1}{\nolinkurl{http://ascl.net/#1}}}
\providecommand{\doarXiv}[1]{\href{https://arxiv.org/abs/#1}{\nolinkurl{https://arxiv.org/abs/#1}}}

\bibitem[{{Abril-Pla} {et~al.}(2023){Abril-Pla}, Andreani, Carroll, Dong, Fonnesbeck, Kochurov, Kumar, Lao, Luhmann, Martin, Osthege, Vieira, Wiecki, \& Zinkov}]{pymc2023}
{Abril-Pla}, O., Andreani, V., Carroll, C., {et~al.} 2023, PeerJ Computer Science, 9, e1516, \dodoi{10.7717/peerj-cs.1516}

\bibitem[{Angus \& McGaugh(2007)}]{Angus2007}
Angus, G.~W., \& McGaugh, S.~S. 2007, Monthly Notices of the Royal Astronomical Society, 383, 417–423, \dodoi{10.1111/j.1365-2966.2007.12403.x}

\bibitem[{Bradley {et~al.}(2026)Bradley, Sip{\H{o}}cz, Robitaille, Tollerud, Deil, Barbary, \& et~al.}]{photutils}
Bradley, L., Sip{\H{o}}cz, B., Robitaille, T., {et~al.} 2026, astropy/photutils: Public Release, v1.13.0,  Zenodo, \dodoi{10.5281/zenodo.596036}

\bibitem[{{Broadhurst} {et~al.}(2025){Broadhurst}, {Li}, {Alfred}, {Diego}, {Morilla}, {Kelly}, {Sun}, {Oguri}, {Williams}, {Windhorst}, {Zitrin}, {Abe}, {Chen}, {Dai}, {Fudamoto}, {Kawai}, {Lim}, {Liu}, {Meena}, {Palencia}, {Smoot}, \& {Williams}}]{Broadhurst2025}
{Broadhurst}, T., {Li}, S.~K., {Alfred}, A., {et~al.} 2025, \apjl, 978, L5, \dodoi{10.3847/2041-8213/ad9aa8}

\bibitem[{{Bushouse} {et~al.}(2017){Bushouse}, {Droettboom}, \& {Greenfield}}]{Bushouse2017}
{Bushouse}, H., {Droettboom}, M., \& {Greenfield}, P. 2017, in Astronomical Society of the Pacific Conference Series, Vol. 512, Astronomical Data Analysis Software and Systems XXV, ed. N.~P.~F. {Lorente}, K.~{Shortridge}, \& R.~{Wayth}, 355

\bibitem[{{Coe} {et~al.}(2012){Coe}, {Umetsu}, {Zitrin}, {Donahue}, {Medezinski}, {Postman}, {Carrasco}, {Anguita}, {Geller}, {Rines}, {Diaferio}, {Kurtz}, {Bradley}, {Koekemoer}, {Zheng}, {Nonino}, {Molino}, {Mahdavi}, {Lemze}, {Infante}, {Ogaz}, {Melchior}, {Host}, {Ford}, {Grillo}, {Rosati}, {Jim{\'e}nez-Teja}, {Moustakas}, {Broadhurst}, {Ascaso}, {Lahav}, {Bartelmann}, {Ben{\'\i}tez}, {Bouwens}, {Graur}, {Graves}, {Jha}, {Jouvel}, {Kelson}, {Moustakas}, {Maoz}, {Meneghetti}, {Merten}, {Riess}, {Rodney}, \& {Seitz}}]{Coe2012}
{Coe}, D., {Umetsu}, K., {Zitrin}, A., {et~al.} 2012, \apj, 757, 22, \dodoi{10.1088/0004-637X/757/1/22}

\bibitem[{Croon {et~al.}(2026)Croon, Crossey, Diego, Kavanagh, \& Palencia}]{Croon2026}
Croon, D., Crossey, B., Diego, J.~M., Kavanagh, B.~J., \& Palencia, J.~M. 2026, Physical Review D, 113, \dodoi{10.1103/4g7h-55r2}

\bibitem[{{Dai} {et~al.}(2018){Dai}, {Venumadhav}, {Kaurov}, \& {Miralda-Escud}}]{Dai2018}
{Dai}, L., {Venumadhav}, T., {Kaurov}, A.~A., \& {Miralda-Escud}, J. 2018, \apj, 867, 24, \dodoi{10.3847/1538-4357/aae478}

\bibitem[{{Diego} {et~al.}(2018){Diego}, {Kaiser}, {Broadhurst}, {Kelly}, {Rodney}, {Morishita}, {Oguri}, {Ross}, {Zitrin}, {Jauzac}, {Richard}, {Williams}, {Vega-Ferrero}, {Frye}, \& {Filippenko}}]{Diego2018}
{Diego}, J.~M., {Kaiser}, N., {Broadhurst}, T., {et~al.} 2018, \apj, 857, 25, \dodoi{10.3847/1538-4357/aab617}

\bibitem[{Diego {et~al.}(2024)Diego, Kei~Li, Amruth, Meena, Broadhurst, Kelly, Filippenko, Williams, Zitrin, Harris, Reina-Campos, Giocoli, Dai, Struble, Treu, Fudamoto, Gilman, Koekemoer, Lim, Palencia, Sun, \& Windhorst}]{Diego2024}
Diego, J.~M., Kei~Li, S., Amruth, A., {et~al.} 2024, Astronomy \&; Astrophysics, 689, A167, \dodoi{10.1051/0004-6361/202450474}

\bibitem[{{D'Silva} {et~al.}(2025){D'Silva}, {Driver}, {Lagos}, {Robotham}, {Adams}, {Conselice}, {Frye}, {Hathi}, {Harvey}, {Koekemoer}, {Ortiz}, {Ricotti}, {Robertson}, {Rutkowski}, {Silver}, {Wilkins}, {Willmer}, {Windhorst}, {Cohen}, {Jansen}, {Summers}, {Coe}, {Grogin}, {Marshall}, {Nonino}, {Pirzkal}, {Ryan}, \& {Yan}}]{DSilva2025}
{D'Silva}, J. C.~J., {Driver}, S.~P., {Lagos}, C. D.~P., {et~al.} 2025, \apj, 990, 44, \dodoi{10.3847/1538-4357/adf19e}

\bibitem[{Fudamoto {et~al.}(2025)Fudamoto, Sun, Diego, Dai, Oguri, Zitrin, Zackrisson, Jauzac, Lagattuta, Egami, Iani, Windhorst, Abe, Bauer, Bian, Bhatawdekar, Broadhurst, Cai, Chen, Chen, Cohen, Conselice, Espada, Foo, Frye, Fujimoto, Furtak, Golubchik, Hsiao, Jolly, Kawai, Kelly, Koekemoer, Kohno, Kokorev, Li, Li, Lin, Magdis, Meena, Niemiec, Nabizadeh, Richard, Steinhardt, Wu, Zhu, \& Zou}]{Fudamoto2025}
Fudamoto, Y., Sun, F., Diego, J.~M., {et~al.} 2025, Nature Astronomy, 9, 428–437, \dodoi{10.1038/s41550-024-02432-3}

\bibitem[{Gabrielli {et~al.}(2024)Gabrielli, Boco, Ghirlanda, Salafia, Salvaterra, Spera, \& Lapi}]{gabrielli2024}
Gabrielli, F., Boco, L., Ghirlanda, G., {et~al.} 2024, Constraining the Initial-Mass Function via Stellar Transients.
\newblock \doarXiv{2409.09118}

\bibitem[{{Gaia Collaboration} {et~al.}(2023){Gaia Collaboration}, {Vallenari}, {Brown}, {Prusti}, {de Bruijne}, {Arenou}, {Babusiaux}, {Biermann}, {Creevey}, {Ducourant}, {Evans}, {Eyer}, {Guerra}, {Hutton}, {Jordi}, {Klioner}, {Lammers}, {Lindegren}, {Luri}, {Mignard}, {Panem}, {Pourbaix}, {Randich}, {Sartoretti}, {Soubiran}, {Tanga}, {Walton}, {Bailer-Jones}, {Bastian}, {Drimmel}, {Jansen}, {Katz}, {Lattanzi}, {van Leeuwen}, {Bakker}, {Cacciari}, {Casta{\~n}eda}, {De Angeli}, {Fabricius}, {Fouesneau}, {Fr{\'e}mat}, {Galluccio}, {Guerrier}, {Heiter}, {Masana}, {Messineo}, {Mowlavi}, {Nicolas}, {Nienartowicz}, {Pailler}, {Panuzzo}, {Riclet}, {Roux}, {Seabroke}, {Sordo}, {Th{\'e}venin}, {Gracia-Abril}, {Portell}, {Teyssier}, {Altmann}, {Andrae}, {Audard}, {Bellas-Velidis}, {Benson}, {Berthier}, {Blomme}, {Burgess}, {Busonero}, {Busso}, {C{\'a}novas}, {Carry}, {Cellino}, {Cheek}, {Clementini}, {Damerdji}, {Davidson}, {de Teodoro}, {Nu{\~n}ez Campos}, {Delchambre}, {Dell'Oro}, {Esquej},
  {Fern{\'a}ndez-Hern{\'a}ndez}, {Fraile}, {Garabato}, {Garc{\'\i}a-Lario}, {Gosset}, {Haigron}, {Halbwachs}, {Hambly}, {Harrison}, {Hern{\'a}ndez}, {Hestroffer}, {Hodgkin}, {Holl}, {Jan{\ss}en}, {Jevardat de Fombelle}, {Jordan}, {Krone-Martins}, {Lanzafame}, {L{\"o}ffler}, {Marchal}, {Marrese}, {Moitinho}, {Muinonen}, {Osborne}, {Pancino}, {Pauwels}, {Recio-Blanco}, {Reyl{\'e}}, {Riello}, {Rimoldini}, {Roegiers}, {Rybizki}, {Sarro}, {Siopis}, {Smith}, {Sozzetti}, {Utrilla}, {van Leeuwen}, {Abbas}, {{\'A}brah{\'a}m}, {Abreu Aramburu}, {Aerts}, {Aguado}, {Ajaj}, {Aldea-Montero}, {Altavilla}, {{\'A}lvarez}, {Alves}, {Anders}, {Anderson}, {Anglada Varela}, {Antoja}, {Baines}, {Baker}, {Balaguer-N{\'u}{\~n}ez}, {Balbinot}, {Balog}, {Barache}, {Barbato}, {Barros}, {Barstow}, {Bartolom{\'e}}, {Bassilana}, {Bauchet}, {Becciani}, {Bellazzini}, {Berihuete}, {Bernet}, {Bertone}, {Bianchi}, {Binnenfeld}, {Blanco-Cuaresma}, {Blazere}, {Boch}, {Bombrun}, {Bossini}, {Bouquillon}, {Bragaglia}, {Bramante}, {Breedt},
  {Bressan}, {Brouillet}, {Brugaletta}, {Bucciarelli}, {Burlacu}, {Butkevich}, {Buzzi}, {Caffau}, {Cancelliere}, {Cantat-Gaudin}, {Carballo}, {Carlucci}, {Carnerero}, {Carrasco}, {Casamiquela}, {Castellani}, {Castro-Ginard}, {Chaoul}, {Charlot}, {Chemin}, {Chiaramida}, {Chiavassa}, {Chornay}, {Comoretto}, {Contursi}, {Cooper}, {Cornez}, {Cowell}, {Crifo}, {Cropper}, {Crosta}, {Crowley}, {Dafonte}, {Dapergolas}, {David}, {David}, {de Laverny}, {De Luise}, \& {De March}}]{Gaia_2023}
{Gaia Collaboration}, {Vallenari}, A., {Brown}, A.~G.~A., {et~al.} 2023, \aap, 674, A1, \dodoi{10.1051/0004-6361/202243940}

\bibitem[{{Gray} \& {Kaur}(2019)}]{Gray2019}
{Gray}, D.~F., \& {Kaur}, T. 2019, \apj, 882, 148, \dodoi{10.3847/1538-4357/ab2fce}

\bibitem[{{Iani} {et~al.}(2023){Iani}, {Rinaldi}, {Caputi}, {Kokorev}, {Annunziatella}, {van Mierlo}, {Kerutt}, {Yang}, {Perez-Gonzalez}, {Costantin}, \& {Caminha}}]{Iani2023}
{Iani}, E., {Rinaldi}, P., {Caputi}, K., {et~al.} 2023, {Unveiling the properties of high-redshift low/intermediate-mass galaxies in Lensing fields with NIRCam Wide Field Slitless Spectroscopy}

\bibitem[{{Ji} \& {Dai}(2025)}]{Ji2025}
{Ji}, L., \& {Dai}, L. 2025, The Astrophysical Journal, 980, 190, \dodoi{10.3847/1538-4357/ada76a}

\bibitem[{{Kaurov} {et~al.}(2019){Kaurov}, {Dai}, {Venumadhav}, {Miralda-Escud{\'e}}, \& {Frye}}]{Kaurov2019}
{Kaurov}, A.~A., {Dai}, L., {Venumadhav}, T., {Miralda-Escud{\'e}}, J., \& {Frye}, B. 2019, \apj, 880, 58, \dodoi{10.3847/1538-4357/ab2888}

\bibitem[{{Levesque} {et~al.}(2005){Levesque}, {Massey}, {Olsen}, {Plez}, {Josselin}, {Maeder}, \& {Meynet}}]{Levesque2005}
{Levesque}, E.~M., {Massey}, P., {Olsen}, K.~A.~G., {et~al.} 2005, \apj, 628, 973, \dodoi{10.1086/430901}

\bibitem[{{Milosavljevi{\'c}} {et~al.}(2007){Milosavljevi{\'c}}, {Koda}, {Nagai}, {Nakar}, \& {Shapiro}}]{Milos2007}
{Milosavljevi{\'c}}, M., {Koda}, J., {Nagai}, D., {Nakar}, E., \& {Shapiro}, P.~R. 2007, \apjl, 661, L131, \dodoi{10.1086/518960}

\bibitem[{{Miralda-Escude}(1991)}]{MiraldaEscude1991}
{Miralda-Escude}, J. 1991, \apj, 379, 94, \dodoi{10.1086/170486}

\bibitem[{{Molnar} {et~al.}(2013){Molnar}, {Broadhurst}, {Umetsu}, {Zitrin}, {Rephaeli}, \& {Shimon}}]{Sandor2013}
{Molnar}, S.~M., {Broadhurst}, T., {Umetsu}, K., {et~al.} 2013, \apj, 774, 70, \dodoi{10.1088/0004-637X/774/1/70}

\bibitem[{{M{\"u}ller} \& {Miralda-Escud{\'e}}(2025)}]{Muller2025}
{M{\"u}ller}, C.~V., \& {Miralda-Escud{\'e}}, J. 2025, \mnras, 536, 1579, \dodoi{10.1093/mnras/stae2652}

\bibitem[{{Niemiec} {et~al.}(2023){Niemiec}, {Jauzac}, {Eckert}, {Lagattuta}, {Sharon}, {Koekemoer}, {Umetsu}, {Acebron}, {Diego}, {Harvey}, {Jullo}, {Kokorev}, {Limousin}, {Mahler}, {Natarajan}, {Nonino}, {Steinhardt}, {Tam}, \& {Zitrin}}]{Niemiec2023}
{Niemiec}, A., {Jauzac}, M., {Eckert}, D., {et~al.} 2023, \mnras, 524, 2883, \dodoi{10.1093/mnras/stad1999}

\bibitem[{{Oguri} {et~al.}(2018){Oguri}, {Diego}, {Kaiser}, {Kelly}, \& {Broadhurst}}]{Oguri2018}
{Oguri}, M., {Diego}, J.~M., {Kaiser}, N., {Kelly}, P.~L., \& {Broadhurst}, T. 2018, \prd, 97, 023518, \dodoi{10.1103/PhysRevD.97.023518}

\bibitem[{{Palencia} {et~al.}(2024){Palencia}, {Diego}, {Kavanagh}, \& {Mart{\'\i}nez-Arrizabalaga}}]{Palencia2024}
{Palencia}, J.~M., {Diego}, J.~M., {Kavanagh}, B.~J., \& {Mart{\'\i}nez-Arrizabalaga}, J. 2024, \aap, 687, A81, \dodoi{10.1051/0004-6361/202347492}

\bibitem[{Palencia {et~al.}(2026)Palencia, Sun, Diego, Fudamoto, Koekemoer, Willmer, Iani, Lin, Pierel, Amruth, Broadhurst, Chen, Dai, Espada, Filippenko, Fujimoto, Kelly, Li, Li, Meena, Miralda-Escudé, Morilla, Struble, Williams, Windhorst, Zackrisson, Zhou, \& Zitrin}]{palencia2026}
Palencia, J.~M., Sun, F., Diego, J.~M., {et~al.} 2026, First Statistical Study of Over 100 Magnified Stellar Events at Redshift $z \approx 0.725$ with JWST.
\newblock \doarXiv{2604.22702}

\bibitem[{{Perrin} {et~al.}(2014){Perrin}, {Sivaramakrishnan}, {Lajoie}, {Elliott}, {Pueyo}, {Ravindranath}, \& {Albert}}]{Perrin2014}
{Perrin}, M.~D., {Sivaramakrishnan}, A., {Lajoie}, C.-P., {et~al.} 2014, in Society of Photo-Optical Instrumentation Engineers (SPIE) Conference Series, Vol. 9143, Space Telescopes and Instrumentation 2014: Optical, Infrared, and Millimeter Wave, ed. J.~M. {Oschmann}, Jr., M.~{Clampin}, G.~G. {Fazio}, \& H.~A. {MacEwen}, 91433X, \dodoi{10.1117/12.2056689}

\bibitem[{{Rigby} {et~al.}(2023){Rigby}, {Perrin}, {McElwain}, {Kimble}, {Friedman}, {Lallo}, {Doyon}, {Feinberg}, {Ferruit}, {Glasse}, {Rieke}, {Rieke}, {Wright}, {Willott}, {Colon}, {Milam}, {Neff}, {Stark}, {Valenti}, {Abell}, {Abney}, {Abul-Huda}, {Acton}, {Adams}, {Adler}, {Aguilar}, {Ahmed}, {Albert}, {Alberts}, {Aldridge}, {Allen}, {Altenburg}, {{\'A}lvarez-M{\'a}rquez}, {Alves de Oliveira}, {Andersen}, {Anderson}, {Anderson}, {Argyriou}, {Armstrong}, {Arribas}, {Artigau}, {Arvai}, {Atkinson}, {Bacon}, {Bair}, {Banks}, {Barrientes}, {Barringer}, {Bartosik}, {Bast}, {Baudoz}, {Beatty}, {Bechtold}, {Beck}, {Bergeron}, {Bergkoetter}, {Bhatawdekar}, {Birkmann}, {Blazek}, {Blome}, {Boccaletti}, {B{\"o}ker}, {Boia}, {Bonaventura}, {Bond}, {Bosley}, {Boucarut}, {Bourque}, {Bouwman}, {Bower}, {Bowers}, {Boyer}, {Bradley}, {Brady}, {Braun}, {Breda}, {Bresnahan}, {Bright}, {Britt}, {Bromenschenkel}, {Brooks}, {Brooks}, {Brown}, {Brown}, {Brown}, {Bunker}, {Burger}, {Bushouse}, {Cale}, {Cameron}, {Cameron},
  {Canipe}, {Caplinger}, {Caputo}, {Cara}, {Carey}, {Carniani}, {Carrasquilla}, {Carruthers}, {Case}, {Catherine}, {Chance}, {Chapman}, {Charlot}, {Charlow}, {Chayer}, {Chen}, {Cherinka}, {Chichester}, {Chilton}, {Chonis}, {Clampin}, {Clark}, {Clark}, {Coe}, {Coleman}, {Comber}, {Comeau}, {Connolly}, {Cooper}, {Cooper}, {Coppock}, {Correnti}, {Cossou}, {Coulais}, {Coyle}, {Cracraft}, {Curti}, {Cuturic}, {Davis}, {Davis}, {Dean}, {DeLisa}, {deMeester}, {Dencheva}, {Dencheva}, {DePasquale}, {Deschenes}, {Hunor Detre}, {Diaz}, {Dicken}, {DiFelice}, {Dillman}, {Dixon}, {Doggett}, {Donaldson}, {Douglas}, {DuPrie}, {Dupuis}, {Durning}, {Easmin}, {Eck}, {Edeani}, {Egami}, {Ehrenwinkler}, {Eisenhamer}, {Eisenhower}, {Elie}, {Elliott}, {Elliott}, {Ellis}, {Engesser}, {Espinoza}, {Etienne}, {Etxaluze}, {Falini}, {Feeney}, {Ferry}, {Filippazzo}, {Fincham}, {Fix}, {Flagey}, {Florian}, {Flynn}, {Fontanella}, {Ford}, {Forshay}, {Fox}, {Franz}, {Fu}, {Fullerton}, {Galkin}, {Galyer}, {Garc{\'\i}a Mar{\'\i}n}, {Gardner},
  {Gardner}, {Garland}, {Garrett}, {Gasman}, {Gaspar}, {Gaudreau}, {Gauthier}, {Geers}, {Geithner}, {Gennaro}, {Giardino}, {Girard}, {Giuliano}, {Glassmire}, \& {Glauser}}]{Rigby2023}
{Rigby}, J., {Perrin}, M., {McElwain}, M., {et~al.} 2023, \pasp, 135, 048001, \dodoi{10.1088/1538-3873/acb293}

\bibitem[{{Robotham} {et~al.}(2023){Robotham}, {D'Silva}, {Windhorst}, {Jansen}, {Summers}, {Driver}, {Wilmer}, \& {Bellstedt}}]{Robotham2023}
{Robotham}, A.~S.~G., {D'Silva}, J.~C.~J., {Windhorst}, R.~A., {et~al.} 2023, \pasp, 135, 085003, \dodoi{10.1088/1538-3873/acea42}

\bibitem[{{Soucail} {et~al.}(1988){Soucail}, {Mellier}, {Fort}, {Mathez}, \& {Cailloux}}]{Soucail1988}
{Soucail}, G., {Mellier}, Y., {Fort}, B., {Mathez}, G., \& {Cailloux}, M. 1988, \aap, 191, L19

\bibitem[{{Stetson}(1987)}]{Stetson1987}
{Stetson}, P.~B. 1987, \pasp, 99, 191, \dodoi{10.1086/131977}

\bibitem[{Sun(2023)}]{sun2023}
Sun, F. 2023, {MAGNIF: Medium-band Astrophysics with the Grism of NIRCam in Frontier Fields}, JWST Proposal ID 2883. Space Telescope Science Institute

\bibitem[{{Venumadhav} {et~al.}(2017){Venumadhav}, {Dai}, \& {Miralda-Escud{\'e}}}]{Venumadhav2017}
{Venumadhav}, T., {Dai}, L., \& {Miralda-Escud{\'e}}, J. 2017, \apj, 850, 49, \dodoi{10.3847/1538-4357/aa9575}

\bibitem[{{Wilde} \& {Frye}(2026)}]{Wilde2026}
{Wilde}, A., \& {Frye}, B. 2026, in American Astronomical Society Meeting Abstracts, Vol. 247, American Astronomical Society Meeting Abstracts, 437.01

\bibitem[{{Williams} {et~al.}(2026){Williams}, {Kelly}, {Windhorst}, {Filippenko}, {Alfred}, {Broadhurst}, {Chen}, {Conselice}, {Cohen}, {Diego}, {Holwerda}, {Koekemoer}, {Li}, {Meena}, {Palencia}, {Ricotti}, {Robertson}, {Sun}, {Yan}, \& {Zitrin}}]{Williams2026}
{Williams}, H., {Kelly}, P.~L., {Windhorst}, R.~A., {et~al.} 2026, \apj, 996, 105, \dodoi{10.3847/1538-4357/ae1966}

\bibitem[{{Windhorst} {et~al.}(2023){Windhorst}, {Cohen}, {Jansen}, {Summers}, {Tompkins}, {Conselice}, {Driver}, {Yan}, {Coe}, {Frye}, {Grogin}, {Koekemoer}, {Marshall}, {O'Brien}, {Pirzkal}, {Robotham}, {Ryan}, {Willmer}, {Carleton}, {Diego}, {Keel}, {Porto}, {Redshaw}, {Scheller}, {Wilkins}, {Willner}, {Zitrin}, {Adams}, {Austin}, {Arendt}, {Beacom}, {Bhatawdekar}, {Bradley}, {Broadhurst}, {Cheng}, {Civano}, {Dai}, {Dole}, {D'Silva}, {Duncan}, {Fazio}, {Ferrami}, {Ferreira}, {Finkelstein}, {Furtak}, {Gim}, {Griffiths}, {Hammel}, {Harrington}, {Hathi}, {Holwerda}, {Honor}, {Huang}, {Hyun}, {Im}, {Joshi}, {Kamieneski}, {Kelly}, {Larson}, {Li}, {Lim}, {Ma}, {Maksym}, {Manzoni}, {Meena}, {Milam}, {Nonino}, {Pascale}, {Petric}, {Pierel}, {Polletta}, {R{\"o}ttgering}, {Rutkowski}, {Smail}, {Straughn}, {Strolger}, {Swirbul}, {Trussler}, {Wang}, {Welch}, {B. Wyithe}, {Yun}, {Zackrisson}, {Zhang}, \& {Zhao}}]{Windhorst_2023}
{Windhorst}, R.~A., {Cohen}, S.~H., {Jansen}, R.~A., {et~al.} 2023, \aj, 165, 13, \dodoi{10.3847/1538-3881/aca163}

\bibitem[{{Yan} {et~al.}(2023){Yan}, {Ma}, {Sun}, {Wang}, {Kelly}, {Diego}, {Cohen}, {Windhorst}, {Jansen}, {Grogin}, {Beacom}, {Conselice}, {Driver}, {Frye}, {Coe}, {Marshall}, {Koekemoer}, {Willmer}, {Robotham}, {D'Silva}, {Summers}, {Nonino}, {Pirzkal}, {Ryan}, {Ortiz}, {Tompkins}, {Bhatawdekar}, {Cheng}, {Zitrin}, \& {Willner}}]{Yan2023}
{Yan}, H., {Ma}, Z., {Sun}, B., {et~al.} 2023, \apjs, 269, 43, \dodoi{10.3847/1538-4365/ad0298}

\end{thebibliography}
\section{Appendix}
\label{sec:appendix}
\begin{figure}[htbp]
    \centering
    \begin{subfigure}[b]{0.48\textwidth}
        \includegraphics[width=\textwidth]{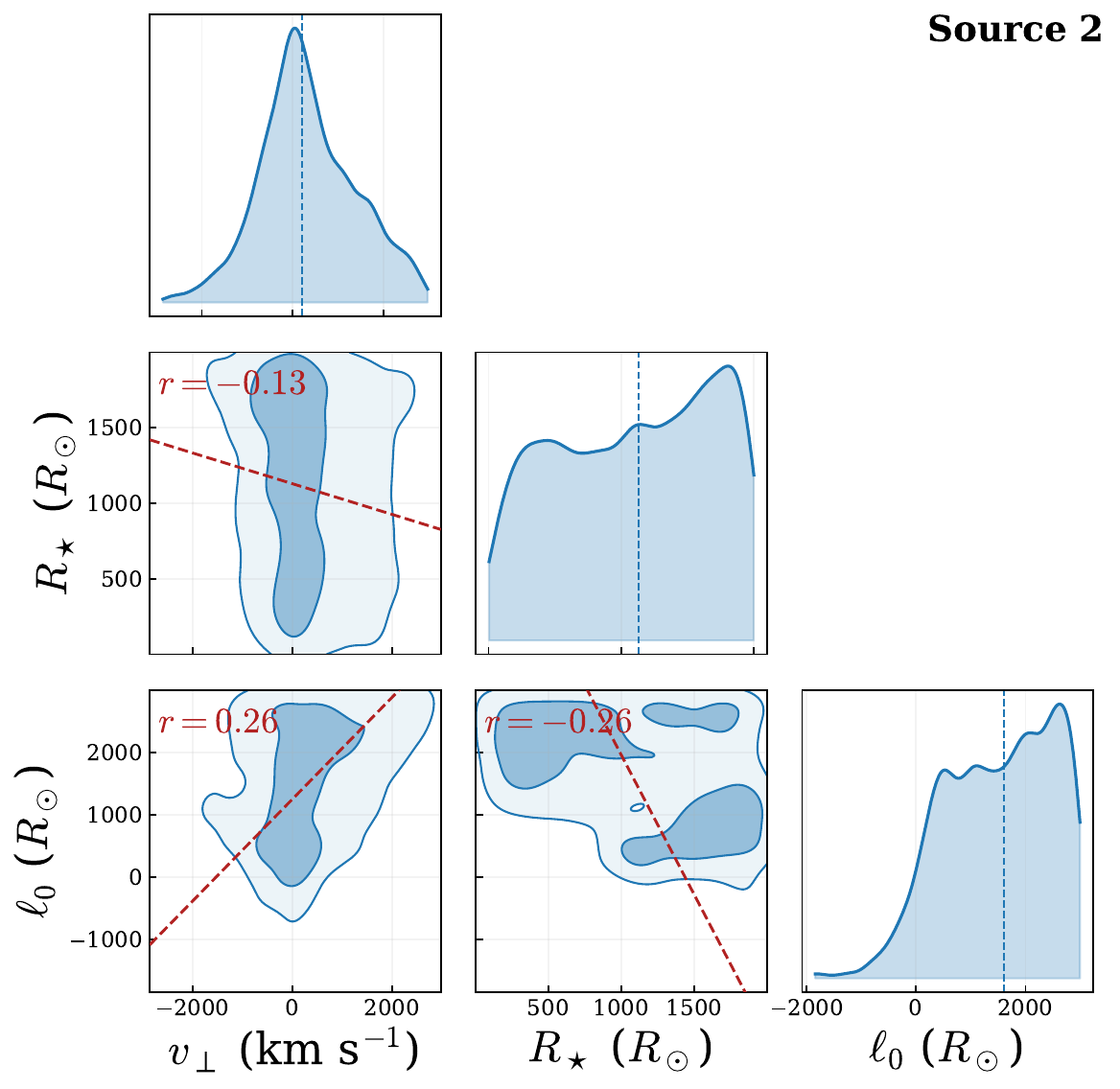}
    \end{subfigure}
    \hfill
    \begin{subfigure}[b]{0.48\textwidth}
        \includegraphics[width=\textwidth]{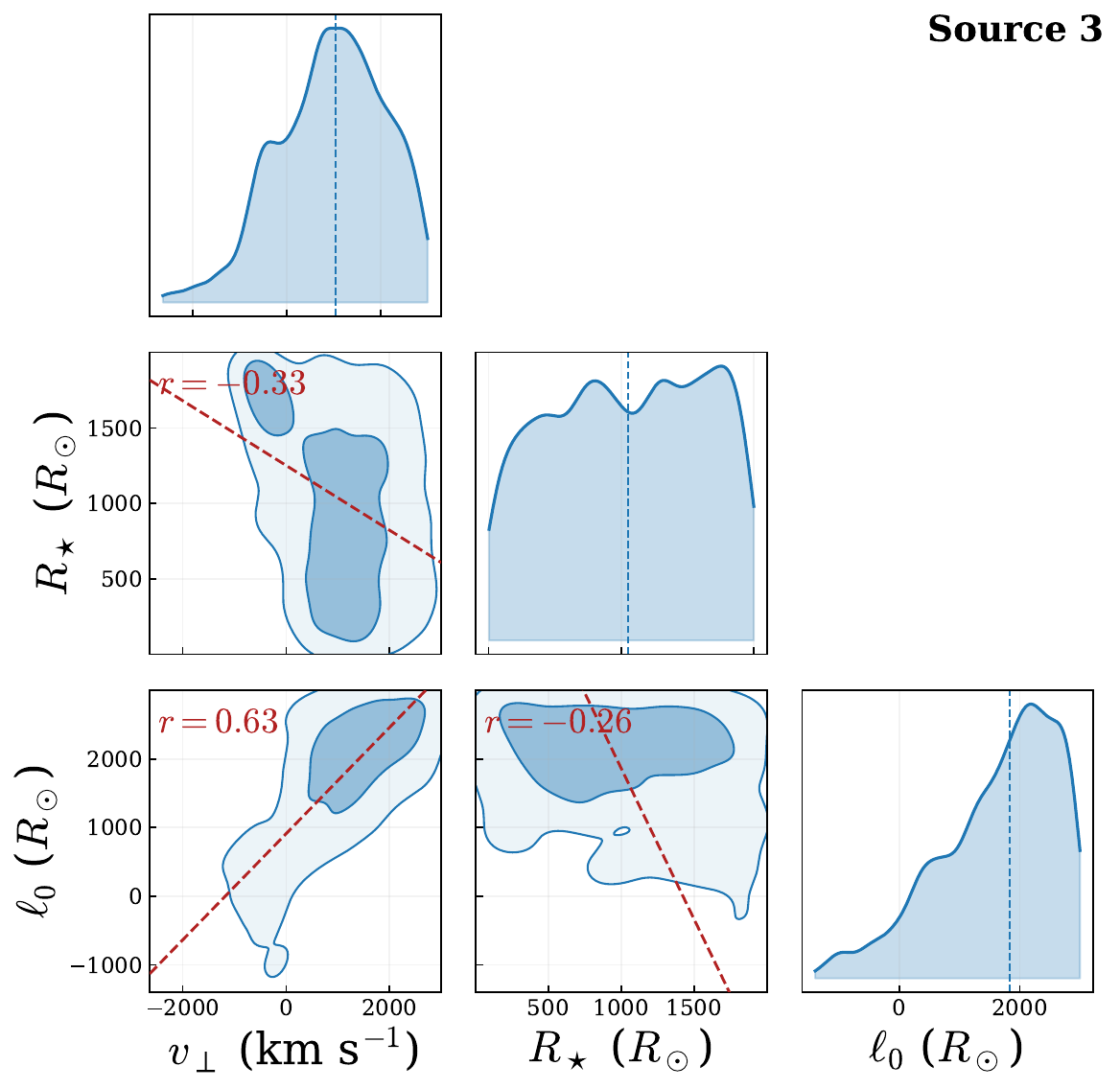}
    \end{subfigure}
    \vspace{0.3cm} 
    \begin{subfigure}[b]{0.48\textwidth}
        \includegraphics[width=\textwidth]{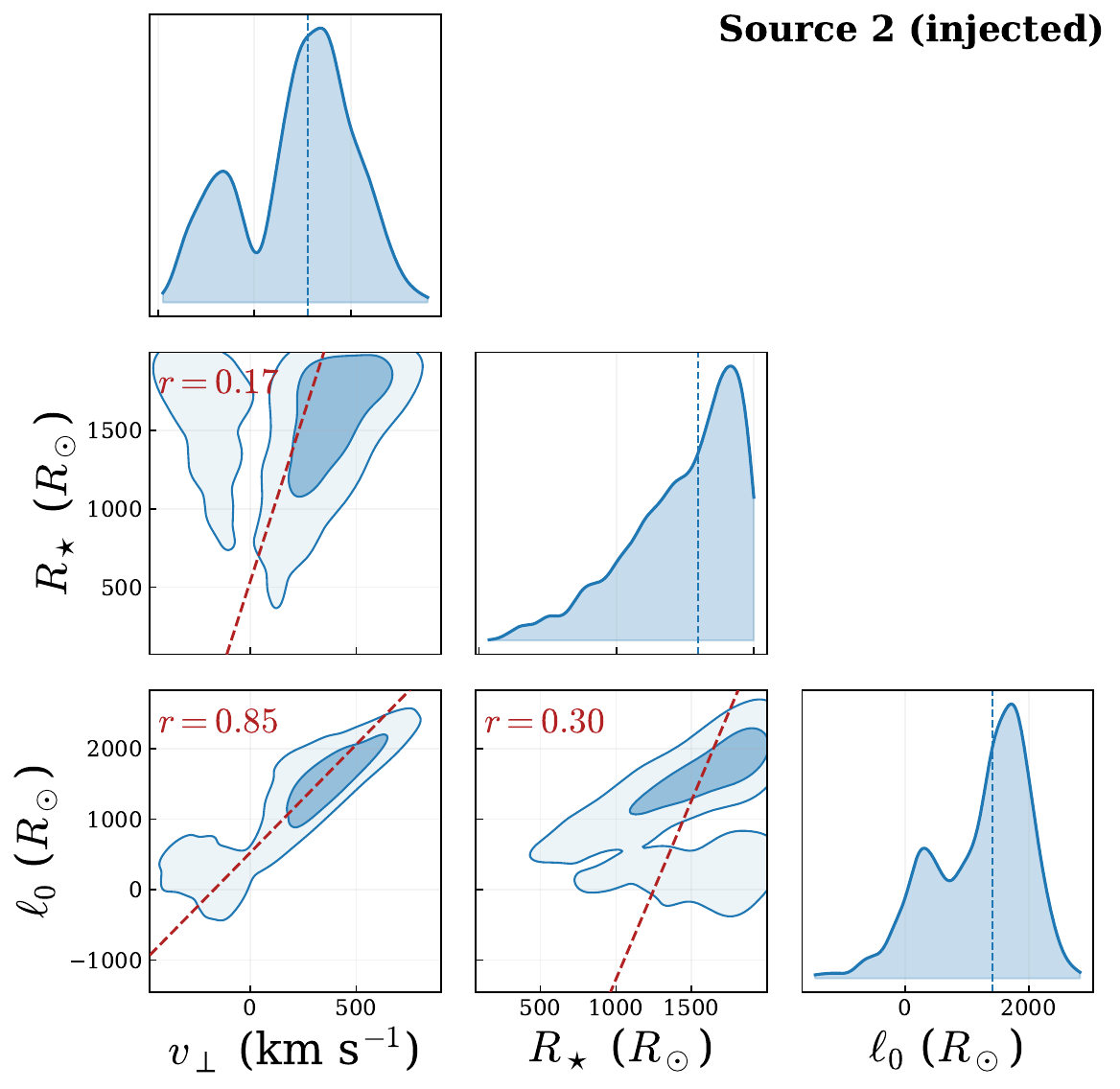}
    \end{subfigure}
    \hfill
    \begin{subfigure}[b]{0.48\textwidth}
        \includegraphics[width=\textwidth]{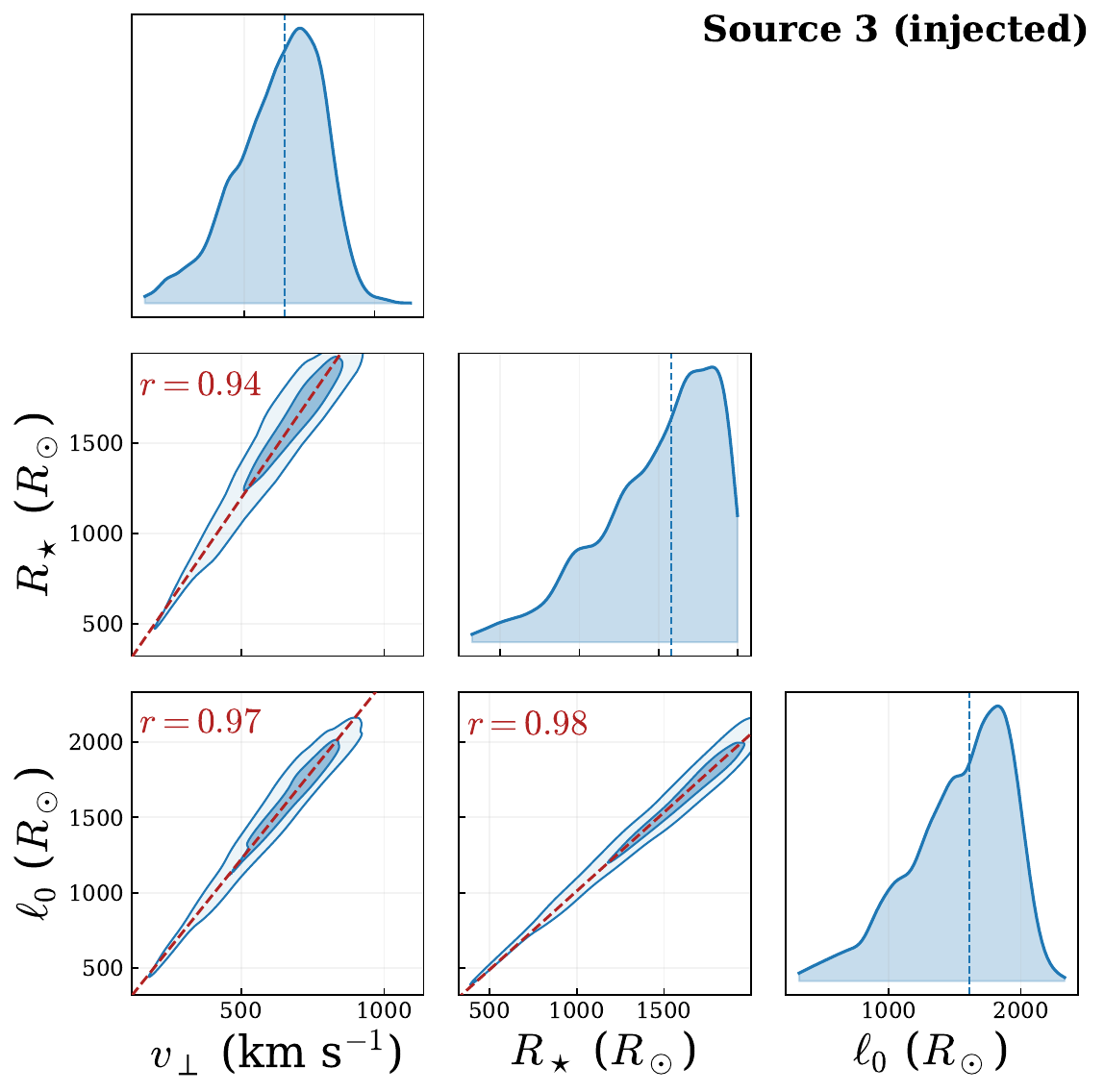}
    \end{subfigure}
    \caption{The same as Figure~\ref{fig:corners}, but for sources 2 and 3.}

\end{figure}

\begin{figure}
    \begin{subfigure}[b]{0.48\textwidth}
        \includegraphics[width=\textwidth]{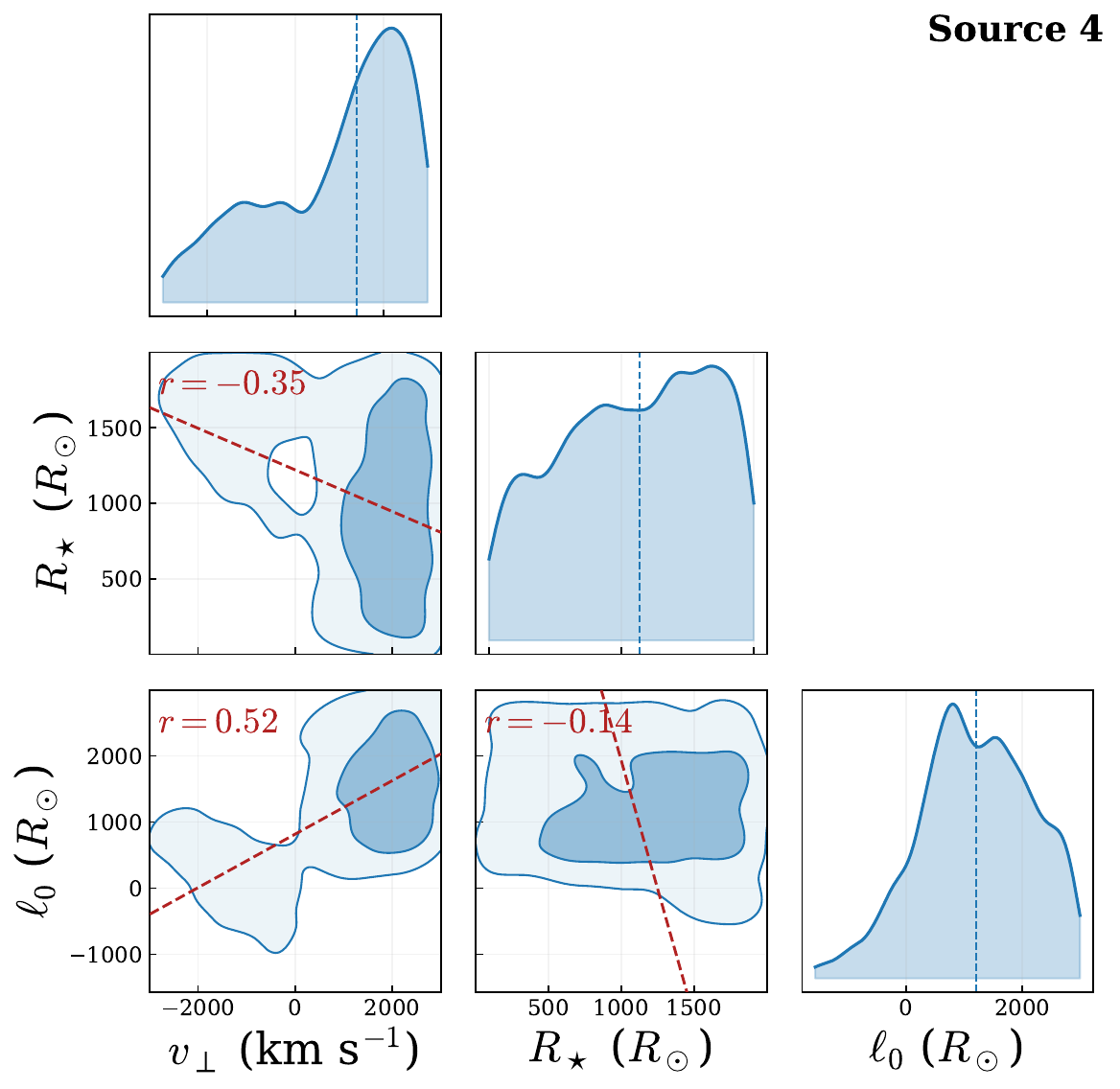}
    \end{subfigure}
    \hfill
    \begin{subfigure}[b]{0.48\textwidth}
        \includegraphics[width=\textwidth]{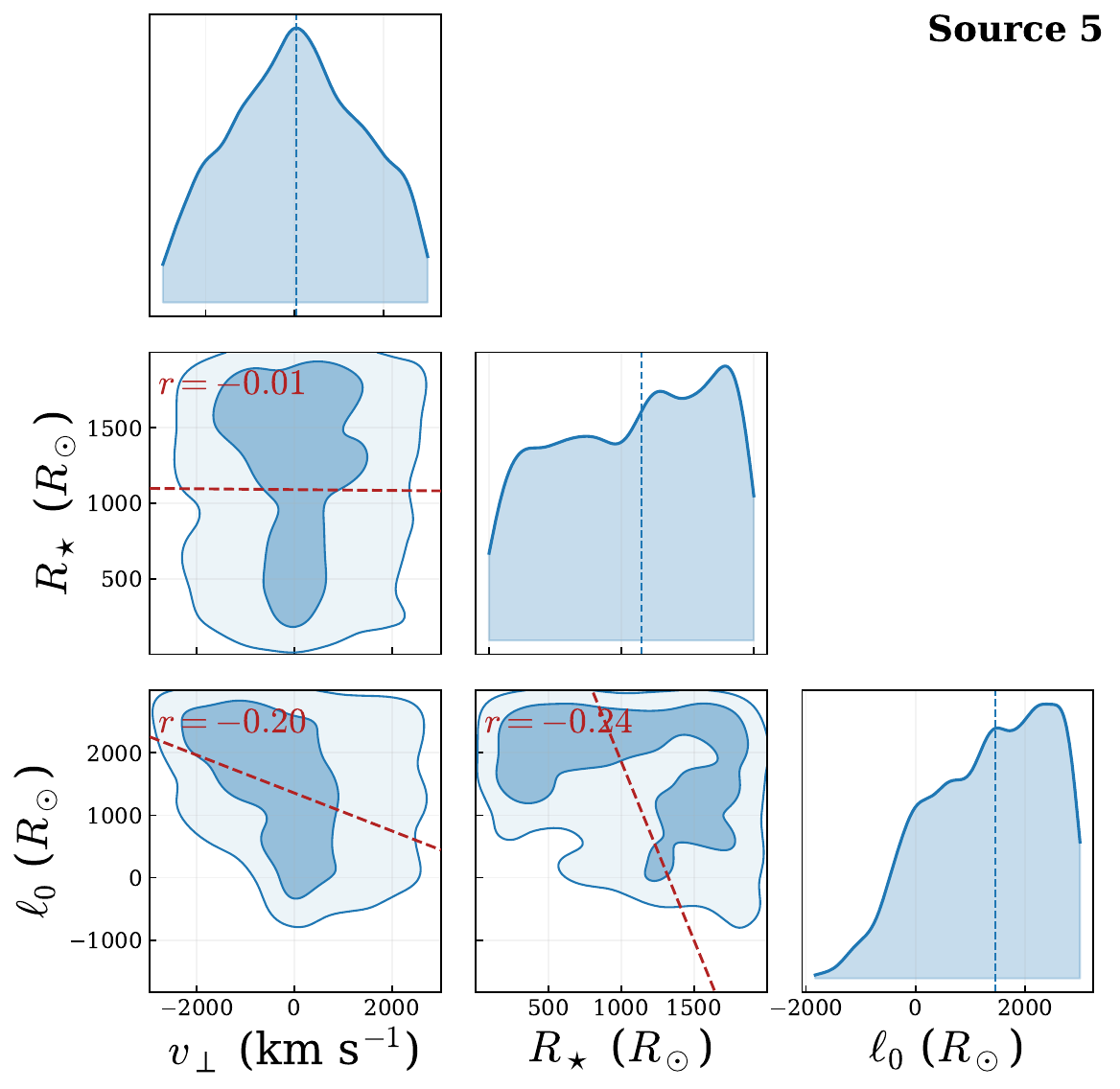}
    \end{subfigure}
    \vspace{0.3cm}
        \begin{subfigure}[b]{0.48\textwidth}
        \includegraphics[width=\textwidth]{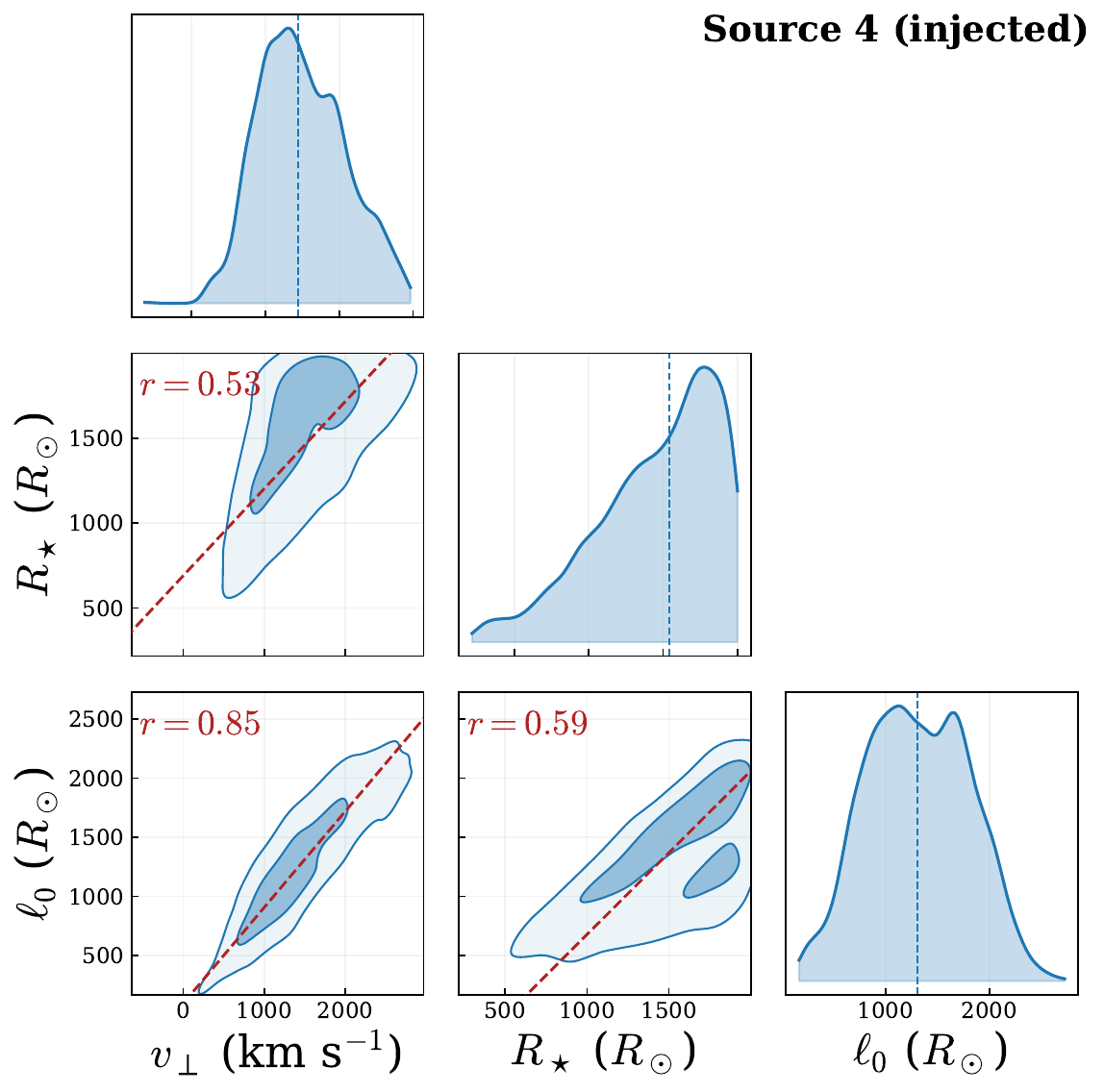}
    \end{subfigure}
    \hfill
    \begin{subfigure}[b]{0.48\textwidth}
        \includegraphics[width=\textwidth]{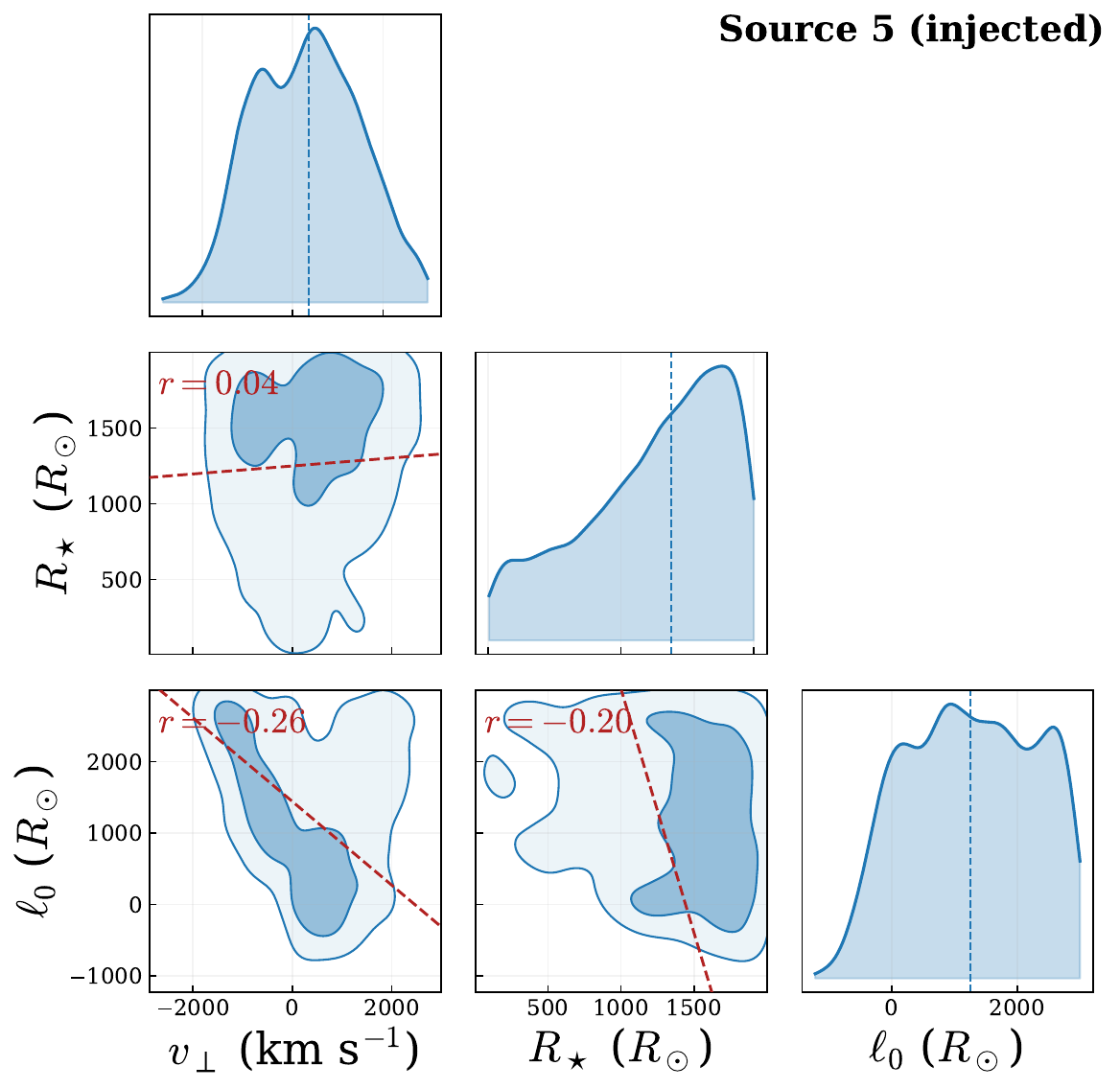}
    \end{subfigure}
    \caption{The same as Figure~\ref{fig:corners}, but for sources 4 and 5.}
\end{figure}

\begin{figure}
    \begin{subfigure}[b]{0.48\textwidth}
        \includegraphics[width=\textwidth]{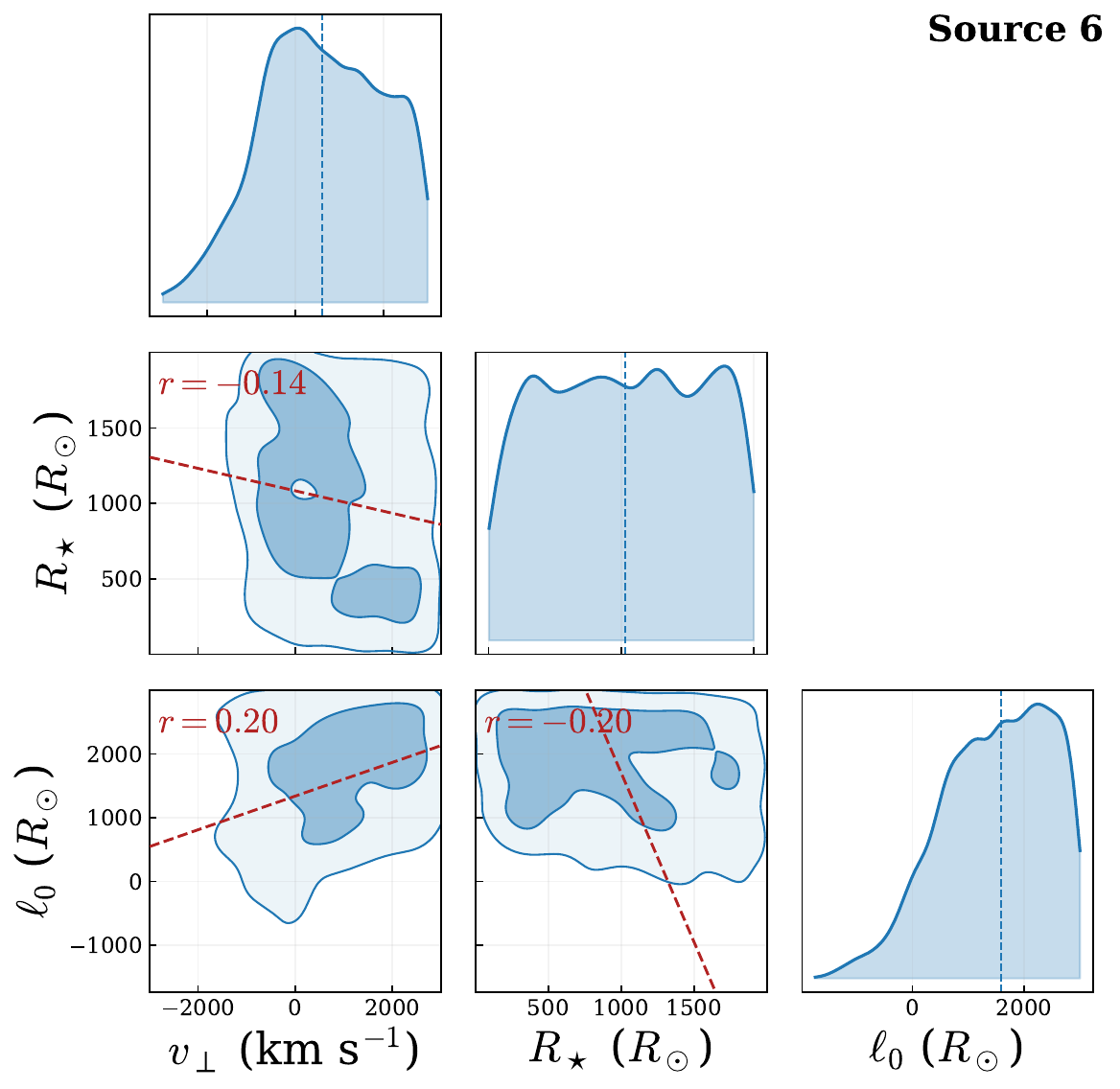}
    \end{subfigure}
    \hfill
    \begin{subfigure}[b]{0.48\textwidth}
        \includegraphics[width=\textwidth]{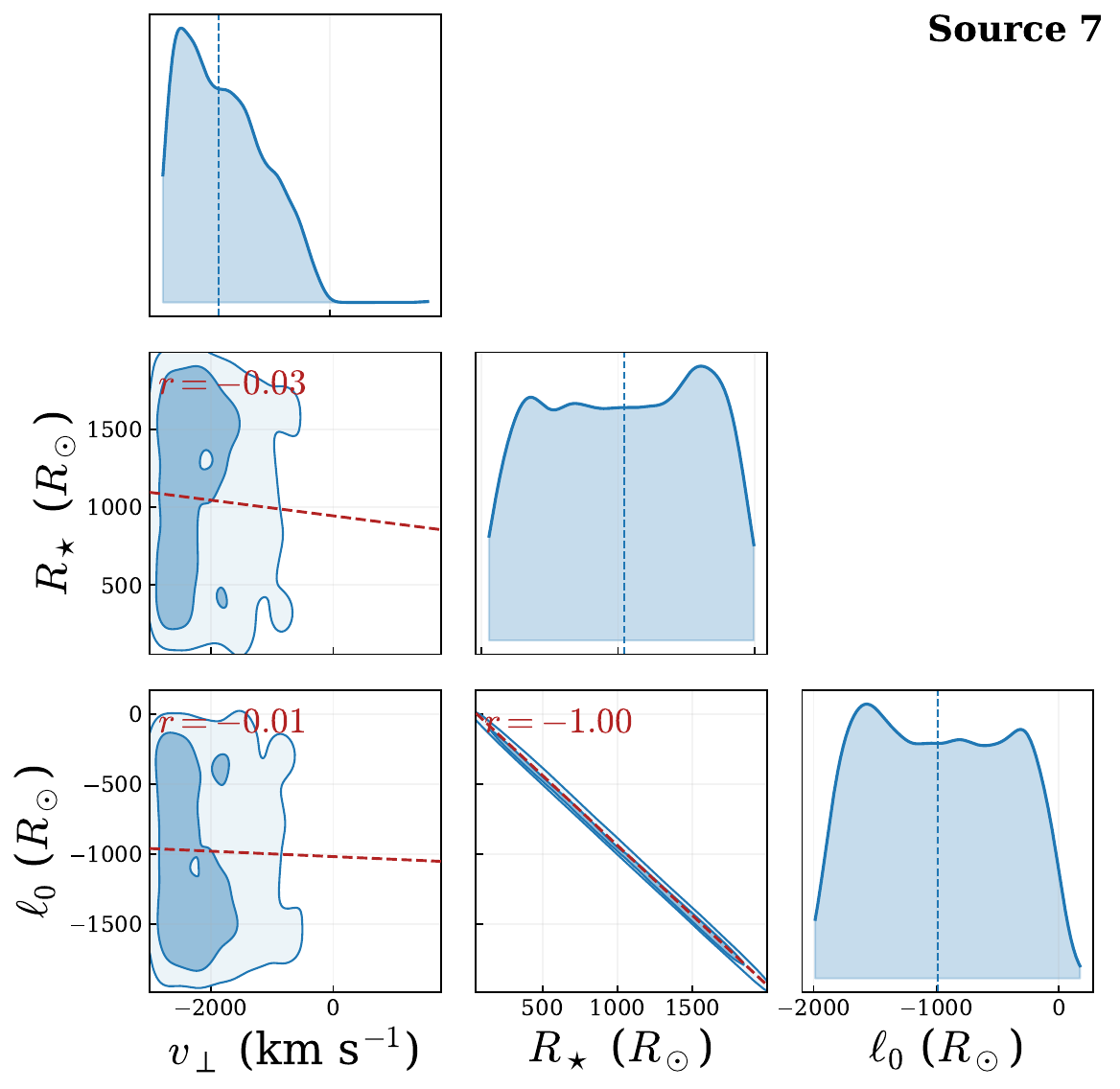}
    \end{subfigure}
    \vspace{0.3cm}
        \begin{subfigure}[b]{0.48\textwidth}
        \includegraphics[width=\textwidth]{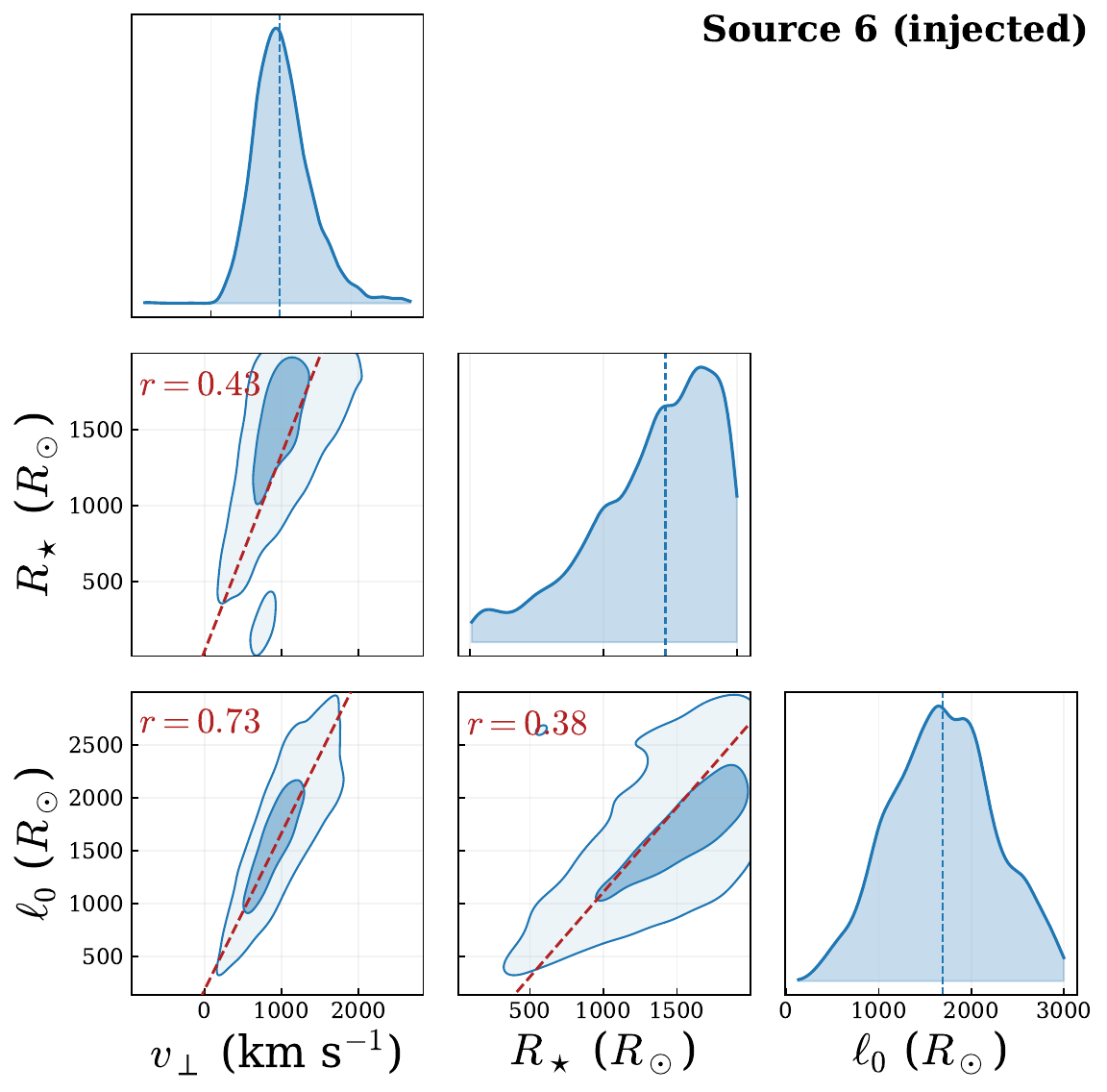}
    \end{subfigure}
    \hfill
    \begin{subfigure}[b]{0.48\textwidth}
        \includegraphics[width=\textwidth]{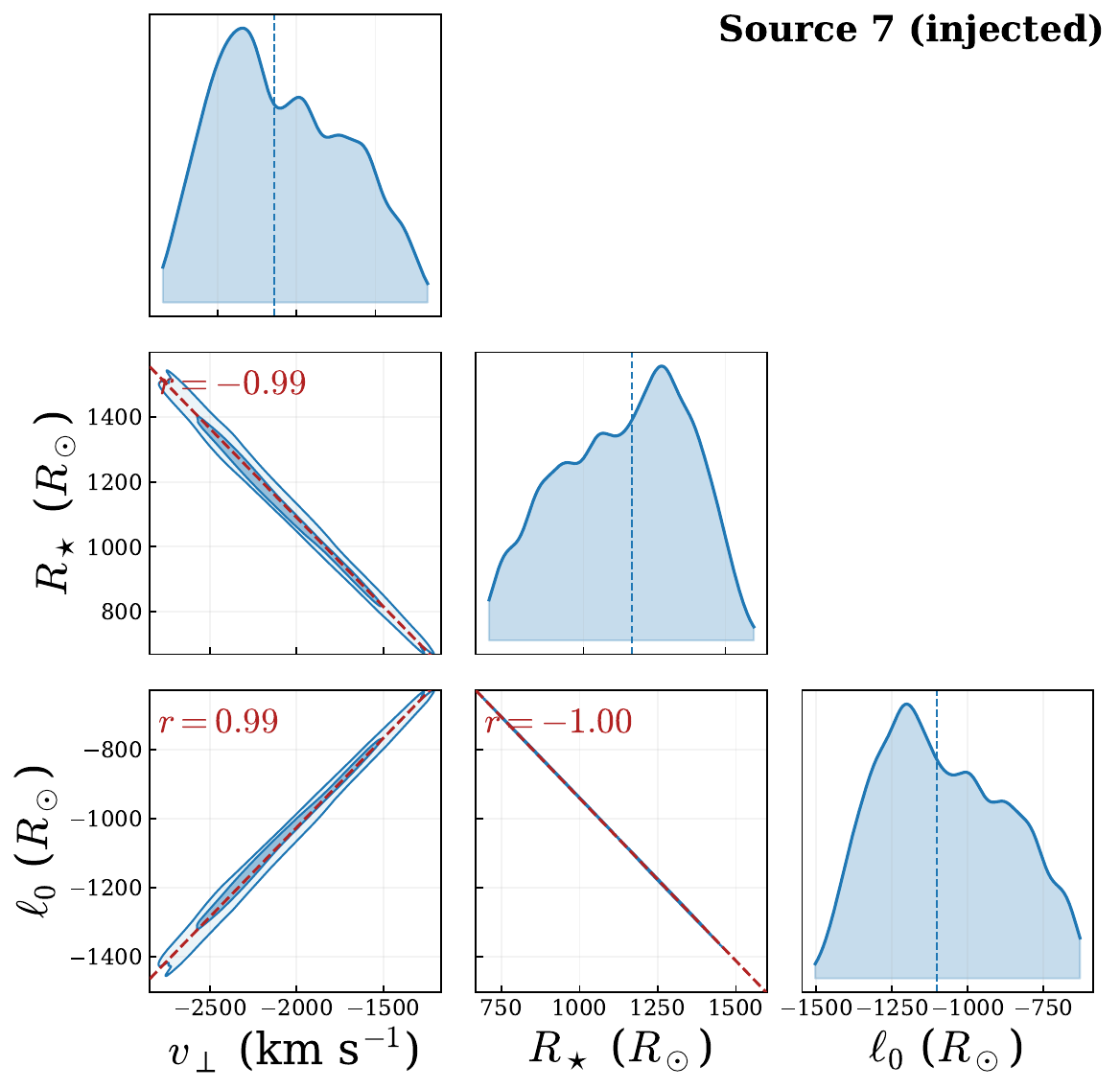}
    \end{subfigure}
    \caption{The same as Figure~\ref{fig:corners}, but for sources 6 and 7.}
\end{figure}

\begin{figure}
    \begin{subfigure}[b]{0.48\textwidth}
        \includegraphics[width=\textwidth]{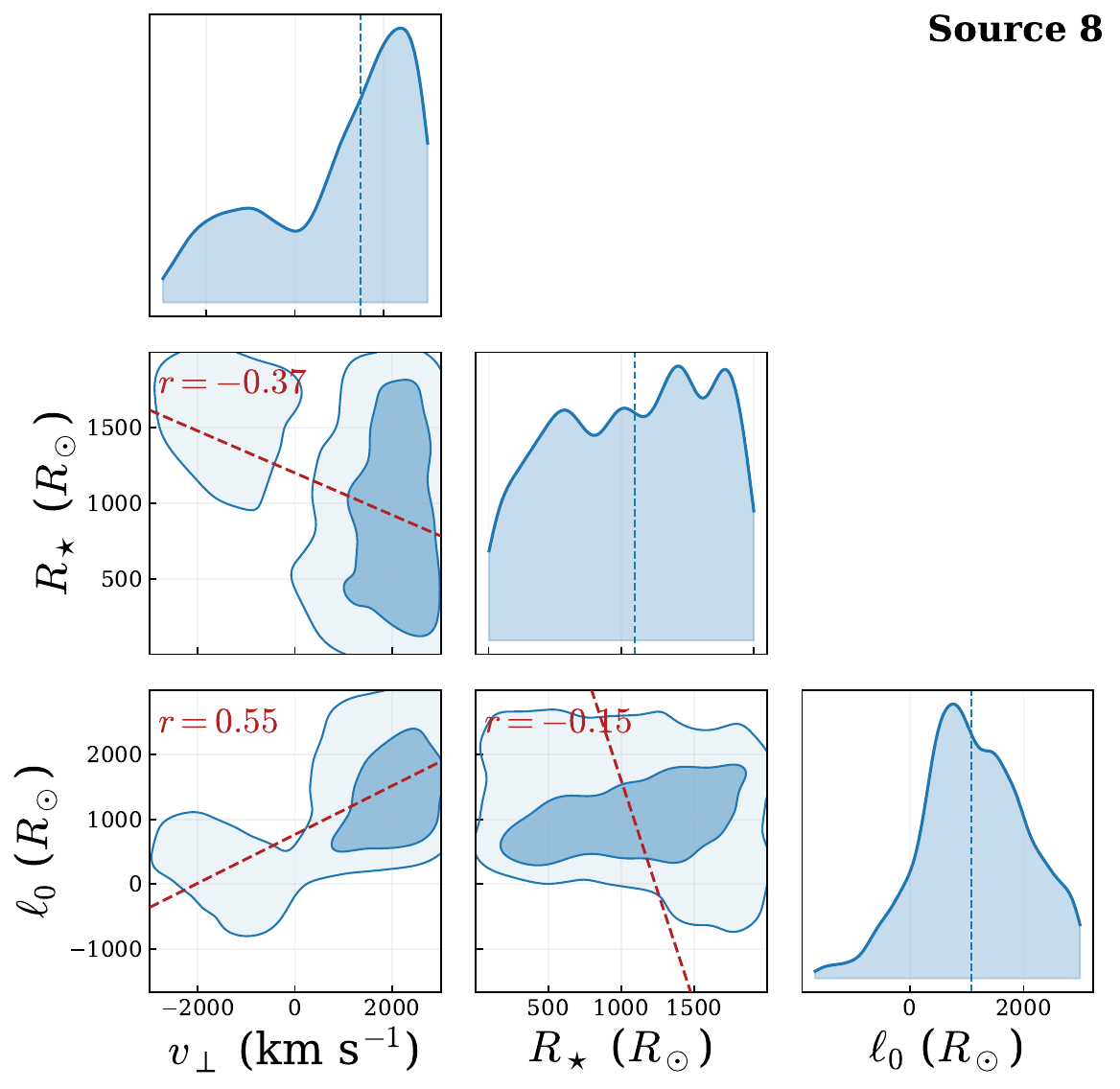}
    \end{subfigure}
    \hfill
    \begin{subfigure}[b]{0.48\textwidth}
        \includegraphics[width=\textwidth]{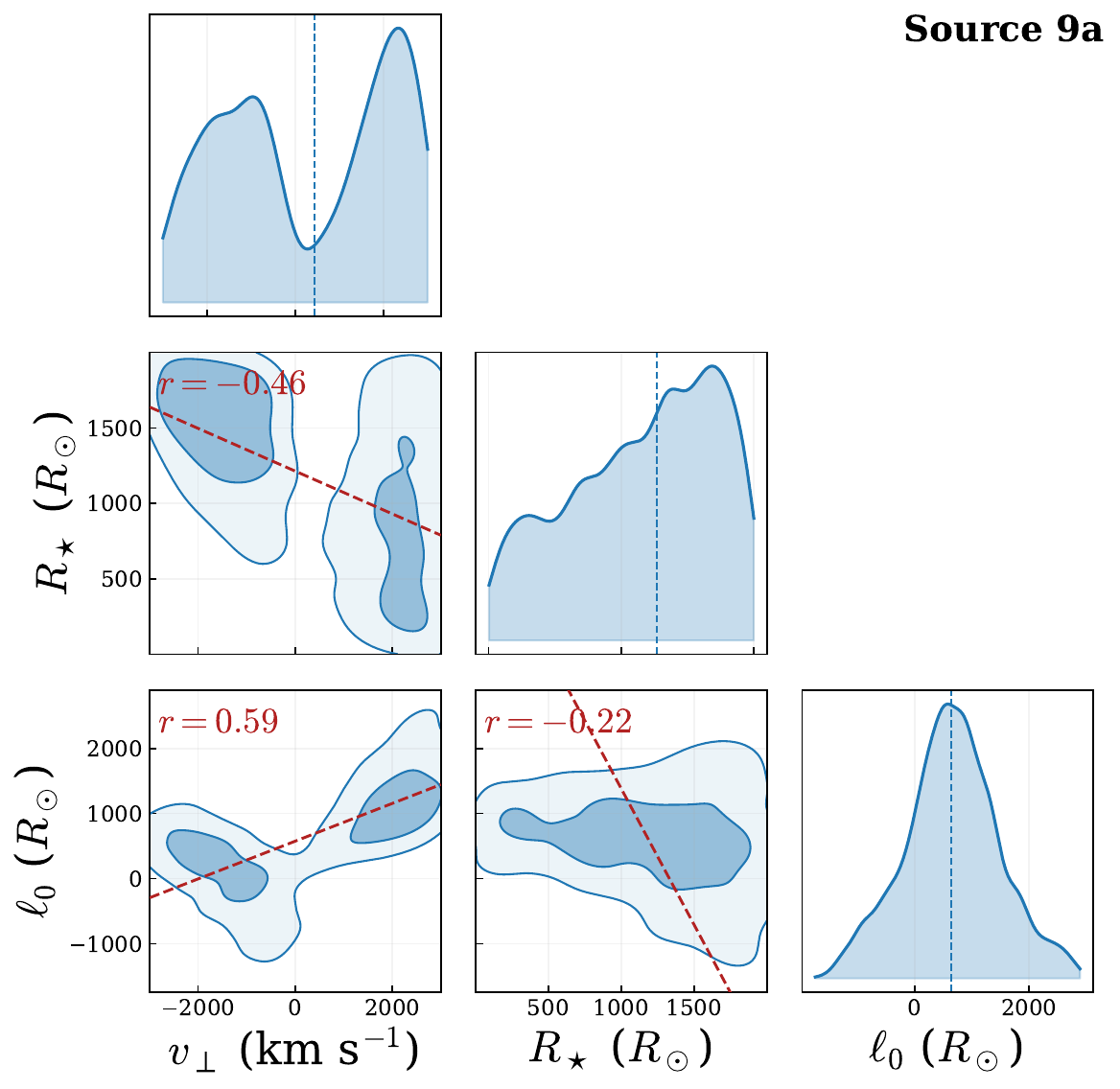}
    \end{subfigure}
    \vspace{0.3cm}
        \begin{subfigure}[b]{0.48\textwidth}
        \includegraphics[width=\textwidth]{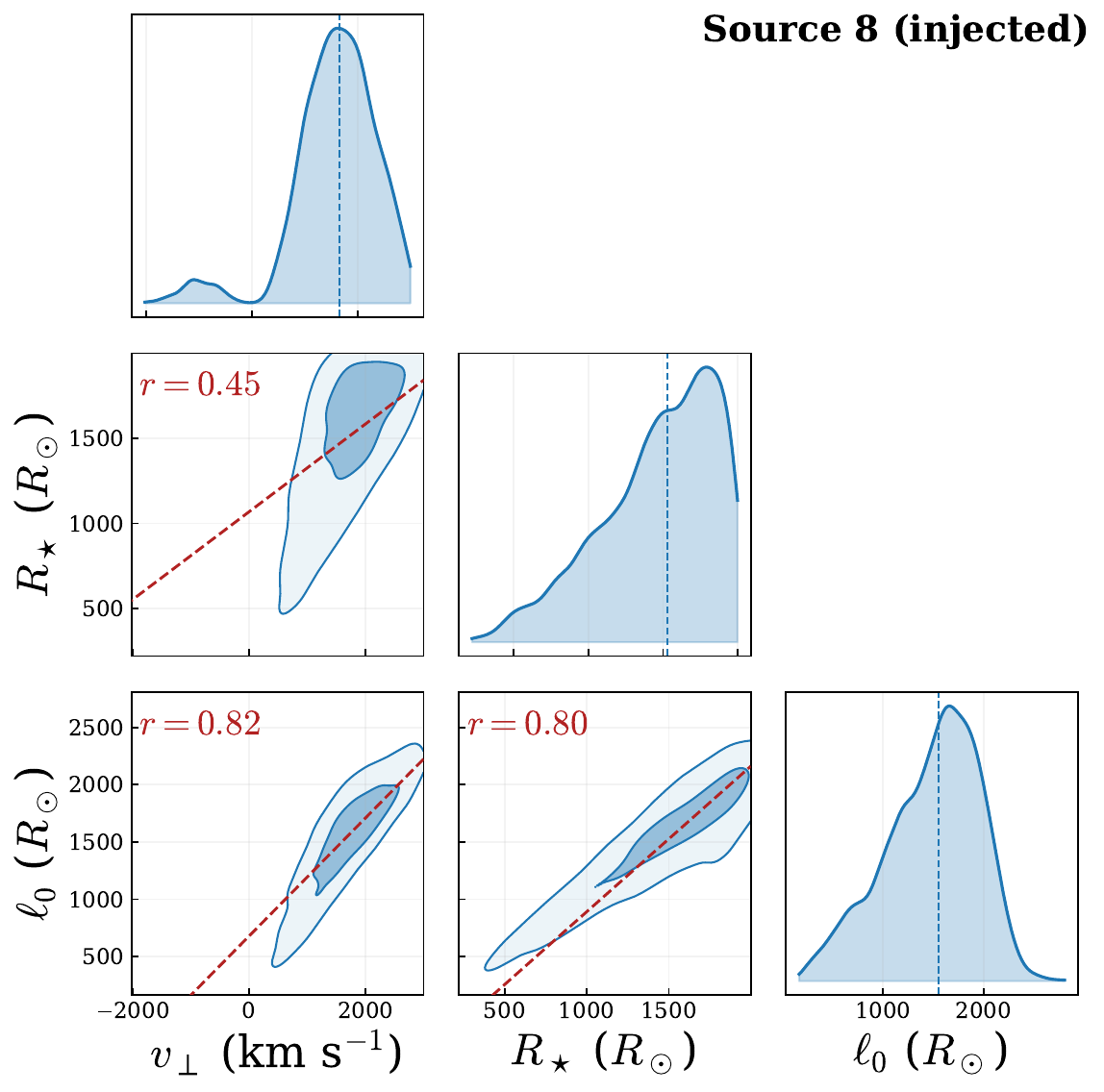}
    \end{subfigure}
    \hfill
    \begin{subfigure}[b]{0.48\textwidth}
        \includegraphics[width=\textwidth]{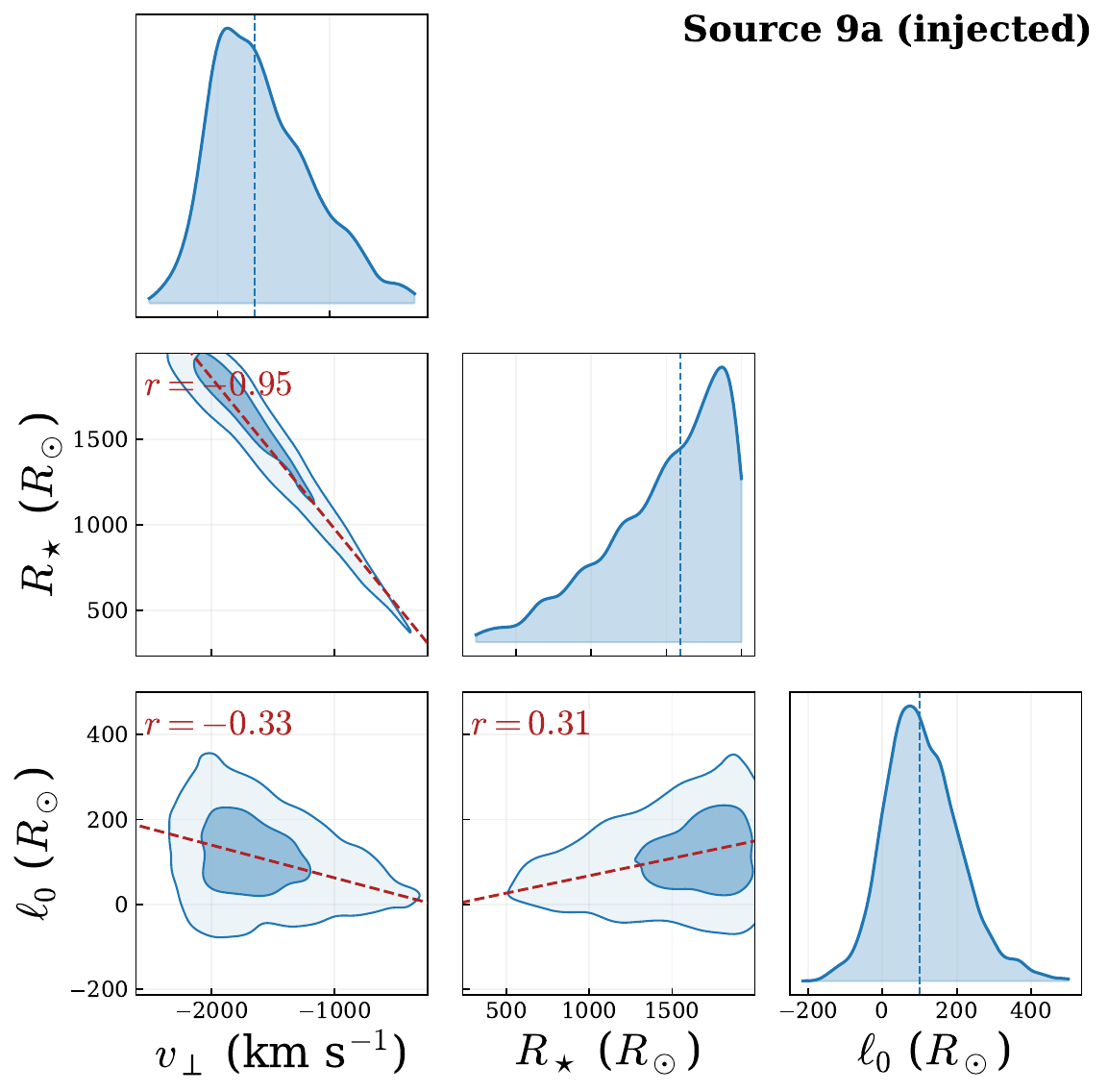}
    \end{subfigure}
    \caption{The same as Figure~\ref{fig:corners}, but for sources 8 and 9a.}
\end{figure}

\begin{figure*}[htbp]
    \centering
    \begin{subfigure}[b]{0.6\textwidth}
        \centering
        \includegraphics[width=\linewidth]{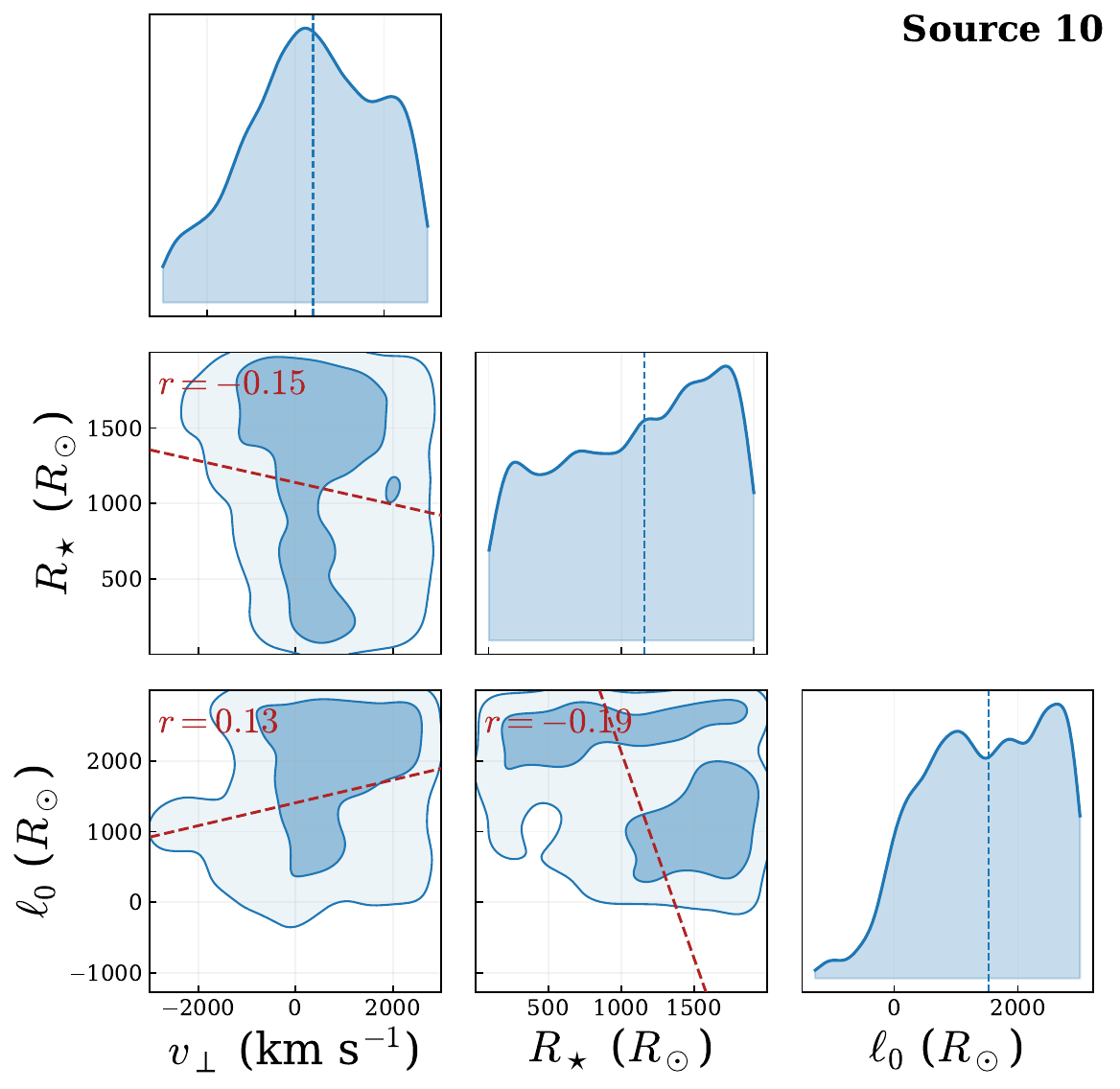}
    \end{subfigure}
    \par\bigskip
    \begin{subfigure}[b]{0.6\textwidth}
        \centering
        \includegraphics[width=\linewidth]{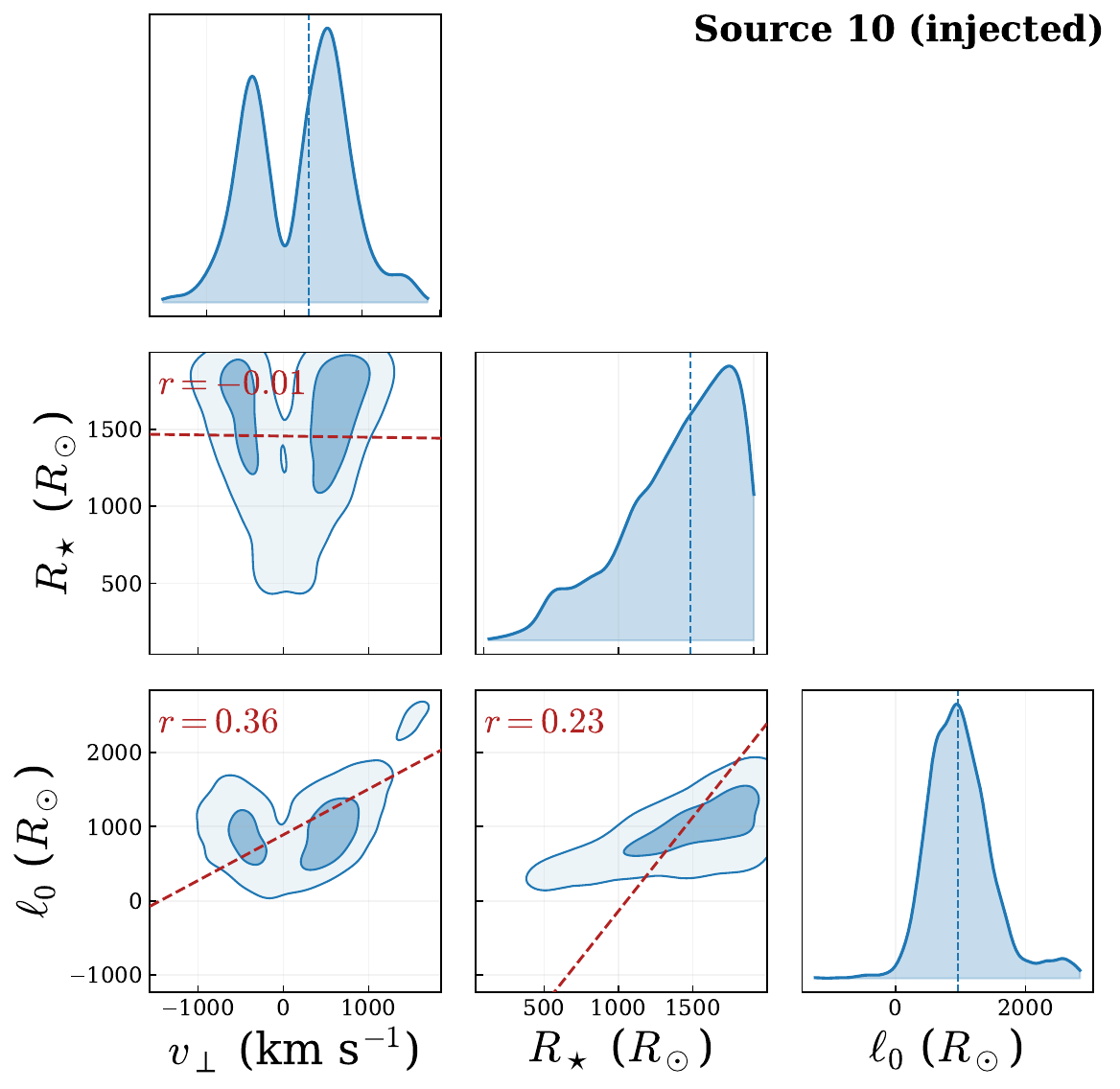}
    \end{subfigure}
    \caption{The same as Figure~\ref{fig:corners}, but for source 10.}
\end{figure*}

\begin{deluxetable}{ccccccccc}
\tablecaption{Best-fit physical parameters recovered from joint F182M+F210M
light-curve fits for caustic-crossing sources in the Dragon arc.}
\label{tab:fit_results}
\tablehead{
\colhead{Source ID} & 
\colhead{$v_\perp$ (km s$^{-1}$)} & 
\colhead{$v_{\perp,inj}$ (km s$^{-1}$)} & 
\colhead{$R$ ($R_\odot$)} & 
\colhead{$R_{inj}$ ($R_{\odot}$)} & 
\colhead{$\ell_0$ ($R_\odot$)} & 
\colhead{$\ell_{0,inj}$ ($R_\odot$)} & 
\colhead{$\chi_{red}^2$} & 
\colhead{$\chi^2_{red,inj}$}
}
\startdata
1 & $-2438^{+4169}_{-0}$  & $2330^{+0}_{-1077}$  & $619^{+1128}_{-329}$  & $1957^{+0}_{-904}$ & $1463^{+0}_{-2676}$ & $-1065^{+497}_{-0}$ & $1.53$ & $1.25$ \\
2 & $144^{+1304}_{-757}$  & $311^{+201}_{-457}$  & $1960^{+0}_{-1563}$  & $1319^{+565}_{-258}$ & $1991^{+625}_{-1559}$ & $1327^{+615}_{-1020}$ & $0.98$ & $0.29$ \\
3 & $827.8^{+1314}_{-1182}$  & $715^{+81}_{-257}$  & $1915^{+0}_{-1539}$  & $1715^{+156}_{-613}$ & $1904^{+706}_{-1338}$ & $1731^{+175}_{-608}$ & $0.76$ & $0.86$  \\
4 & $2735^{+0}_{-3818}$  & $1593^{+539}_{-680}$  & $1574^{+160}_{-1140}$  & $1634^{+233}_{-599}$ & $1548^{+666}_{-1264}$ & $1619^{+231}_{-608}$ & $5.32$ & $1.80$ \\
5 & $-1631^{+3326}_{-0}$  & $664^{+866}_{-1577}$  & $1904^{+0}_{-1500}$  & $1730^{+84}_{-1120}$ & $1913^{+618}_{-1828}$ & $144^{+2293}_{-10}$ & $1.38$ & $1.01$ \\
6 & $1123^{+1059}_{-1840}$  & $860^{+555}_{-212}$  & $1880^{+0}_{-1529}$  & $1701^{+144}_{-800}$ & $1858^{+680}_{-1342}$ & $1698^{+575}_{-593}$ & $3.37$ & $0.84$ \\
7 & $-2797^{+1683}_{-0}$  & $-1877^{+252}_{-636}$  & $1350^{+335}_{-962}$  & $1014^{+364}_{-127}$ & $-1256^{+942}_{-365}$ & $-955^{+122}_{-341}$ & $5.39$ & $1.33$ \\
8 & $2296^{+300}_{-3497}$  & $2272^{+61}_{-1286}$  & $1485^{+247}_{-1083}$  & $1850^{+17}_{-786}$ & $1475^{+641}_{-1200}$ & $1944^{+12}_{-922}$ & $1.39$ & $1.22$ \\
9a & $-1923^{+4364}_{-0}$  & $-1977^{+841}_{-45}$  & $1493^{+268}_{-970}$  & $1917^{+0}_{-837}$ & $342^{+1091}_{-476}$ & 
$97^{+106}_{-81}$ & $1.02$ & $0.96$ \\
10 & $2943^{+0}_{-4120}$  & $-313^{+1082}_{-176}$  & $26^{+1750}_{-0}$  & $1108^{+753}_{-85}$ & $-21^{+2633}_{-0}$ & 
$693^{+745}_{-142}$ & $2.73$ & $1.33$ \\
\\
\enddata
\tablecomments{The ``\textit{inj}'' flag denotes fit-parameters after observation dates are injected. After injected observation dates, errors for each parameter are reduced significantly. Reduced chi-squared values before and after injection dates are included to demonstrate best-fit improvement with additional observations.}
\end{deluxetable}

\end{document}